\documentclass[a4paper]{article}

\usepackage{a4wide}
\usepackage[UKenglish]{babel}
\usepackage{graphicx,subfigure,pgfplots}
	\pgfplotsset{compat=newest}
	\usepgfplotslibrary{fillbetween}
	\usepgfplotslibrary{groupplots}
	\usetikzlibrary{patterns}
	\usetikzlibrary{matrix}
	\usetikzlibrary{positioning}
\usepackage{epstopdf}
	\graphicspath{{./}{figures/test_Boltzmann/}{figures/test_Boltzmann_D0/}{figures/test_Boltzmann_Enskog/}{figures/test_Boltzmann_Enskog_D0/}{figures/test_macro/}{figures/test_omo/}}
\usepackage[colorlinks=true,linkcolor=blue,citecolor=red]{hyperref}
\usepackage{xcolor}
\usepackage{amsfonts,amsmath,amssymb,amsthm,mathtools,bbold,mathrsfs}
\usepackage[inline]{enumitem}
\usepackage{algorithm,algorithmic}
\usepackage{diagbox}

\newtheorem{theorem}{Theorem}[section]

\theoremstyle{remark}\newtheorem{remark}[theorem]{Remark}

\newcommand{\abs}[1]{\left\lvert#1\right\rvert}
\DeclarePairedDelimiter\ave{\langle}{\rangle}

\newcommand{\N}{\mathbb{N}}

\newcommand{\sP}{\mathscr{P}}

\newcommand{\R}{\mathbb{R}}
\renewcommand{\S}{\mathbb{S}}

\allowdisplaybreaks

\begin{document}
\title{The Aw-Rascle traffic model:\\ Enskog-type kinetic derivation and generalisations}

\author{Giacomo Dimarco\thanks{Department of Mathematics and Computer Sciences, University of Ferrara, Via Machiavelli 35, 44121 Ferrara, Italy
            (\texttt{giacomo.dimarco@unife.it})} \and
        Andrea Tosin\thanks{Department of Mathematical Sciences ``G. L. Lagrange'', Politecnico di Torino, Corso
        		Duca degli Abruzzi 24, 10129 Torino, Italy
            (\texttt{andrea.tosin@polito.it})}}
\date{}

\maketitle

\begin{abstract}
We study the derivation of second order macroscopic traffic models from kinetic descriptions. In particular, we recover the celebrated Aw-Rascle model as the hydrodynamic limit of an Enskog-type kinetic equation out of a precise characterisation of the microscopic binary interactions among the vehicles. Unlike other derivations available in the literature, our approach unveils the multiscale physics behind the Aw-Rascle model. This further allows us to generalise it to a new class of second order macroscopic models complying with the Aw-Rascle consistency condition, namely the fact that no wave should travel faster than the mean traffic flow.

\medskip


\medskip


\end{abstract}

\section{Introduction}
The kinetic description of vehicular traffic is probably one of the first examples in which methods of the statistical physics were applied to a particle system which was not a standard gas. Such an approach was initiated by the Russian physicist Ilya Prigogine in the sixties~\cite{herman1979Science,prigogine1961PROC,prigogine1960OR,prigogine1971BOOK}, in an attempt to explain the emergence of collective properties as a result of individual ones in systems composed by human beings instead of molecules. In more recent times, the application of kinetic equations to various systems of interacting agents (see for instance \cite{borsche2012CMA,puppo2016CMS,visconti2017MMS} for application to traffic flows) has gained a lot of momentum. These equations, and more in general the mathematical-physical theory on which they are grounded, have proved to be powerful tools to increase the understanding of multi-agent systems, particularly as far as the exploration of the interconnections among their properties at different scales is concerned~\cite{naldi2010BOOK,pareschi2013BOOK}. 

Coming back to vehicular traffic, in the literature there exists at least two other modelling approaches based on differential equations. On one hand, there are the so-called microscopic models, which describe the vehicles as point particles moving according to prescribed acceleration/deceleration laws. We recall here, in particular, the well known follow-the-leader and optimal velocity models~\cite{bando1995PRE,gazis1961OR}. On the other hand, there are the macroscopic models, which, inspired by the hyperbolic conservation/balance laws of fluid dynamics, treat the vehicles as a continuum with density~\cite{piccoli2009ENCYCLOPEDIA}. In this case, one distinguishes between first order models, which rely on the mass conservation only, cf. e.g.~\cite{cristiani2019M2AS,lighthill1955PRSL,richards1956OR}, and second order models, which include also an equation for the conservation or the balance of the mean speed, cf. e.g.~\cite{payne1971MMPS}. Second order models allow one to overcome the issue of the unbounded acceleration of the vehicles, which first order models may suffer from, see~\cite{laurent-brouty2019PREPRINT} for a very recent contribution on this. However, they may fail to reproduce the correct anisotropy of the interactions among the vehicles, namely the fact that vehicles are mainly influenced by the dynamics ahead than by those behind them. This issue was first pointed out by Daganzo in~\cite{daganzo1995TR} and later solved by Aw and Rascle~\cite{aw2000SIAP} and, independently, by Zhang~\cite{zhang2002TRB}. They proposed a heuristic second order hyperbolic traffic model, whose characteristic speeds never exceed the speed of the flow. In this way, the small disturbances produced by a vehicle propagate more slowly than the vehicles themselves, thereby guaranteeing that the movement of each vehicle affects only the vehicles behind.

An interesting theoretical problem, left largely unexplored in the original papers~\cite{aw2000SIAP,zhang2002TRB}, is the derivation of the Aw-Rascle macroscopic model from first principles. In~\cite{herty2007NHM,klar1997JSP,klar2000SIAP}, the authors were the first to obtain the Aw-Rascle model as a hydrodynamic limit of a kinetic description based on an Enskog-type equation. Their approach is very much inspiring, because it suggests to look at the Enskog-type kinetic description instead of the more classical Boltzmann-type one. On the other hand, in these cited works, the authors did not focus on the explicit characterisation of fundamental microscopic interactions able to generate, at the macroscopic level, the Aw-Rascle model. Moreover, in~\cite{klar1997JSP,klar2000SIAP} the hydrodynamic limit is performed by postulating the existence of an equilibrium kinetic distribution function, which is not exhibited explicitly. In addition, partly heuristic closures of other terms appearing in the equations are used. More recently, also the direct link between follow-the-leader microscopic models and the Aw-Rascle macroscopic model has been explored. In particular, in~\cite{difrancesco2017MBE,difrancesco2015ARMA} the authors prove that the trajectories of the former converge, in the $1$-Wasserstein metric, to the unique entropy solution of the latter when a suitable large particle limit is considered. Their strategy consists in interpreting the follow-the-leader model as a discrete Lagrangian approximation of the target macroscopic model. We observe that this approach, although successful from the analytical point of view, does not explain the actual multiscale physics behind the derivation of the Aw-Rascle model from a microscopic particle model.

In this paper, we investigate the possibility to obtain the Aw-Rascle model as the hydrodynamic limit of kinetic descriptions of the traffic system. The highlights of our study, which differentiate it from the other contributions recalled above, may be summarised as follows:
\begin{itemize}
\item we derive explicitly a minimal set of fundamental features of the microscopic interactions among the vehicles, necessary to generate the Aw-Rascle macroscopic model. Furthemore, we link some key elements of the Aw-Rascle model, such as e.g., the so-called ``traffic pressure'', to precise characteristics of the microscopic interactions. We observe that, in the modelling of multi-agent systems, the microscopic model of the agent behaviour is often largely heuristic and, as such, somewhat arbitrary. In this respect, our result helps to identify a paradigmatic class of interaction rules among the vehicles, which give rise to a physically consistent macroscopic traffic model;
\item we elucidate the multiscale physical structure underlying the Aw-Rascle model. In particular, we show that an Enskog-type kinetic description, as opposed to a Boltzmann-type one, is ultimately necessary to derive it, because the anticipatory nature of the Aw-Rascle dynamics may be understood as the hydrodynamic result of local and non-local microscopic interactions happening on different time scales;
\item taking advantage of the previous analysis, we show how to generalise the Aw-Rascle model to new classes of second order macroscopic traffic models, which take correctly into account the anisotropy of the interactions among the vehicles.
\end{itemize}

In more detail, the paper is organised as follows: in Section~\ref{sect:micro}, we discuss the microscopic interactions at the basis of the whole theory. In Section~\ref{sect:hydro.Boltzmann}, we show that a Boltzmann-type kinetic description does not give rise to the Aw-Rascle model in the hydrodynamic limit nor, more in general, to a macroscopic model correctly reproducing the anisotropy of the vehicle interactions 
Conversely, in Section~\ref{sect:hydro.Enskog}, we prove that the original Aw-Rascle model can be obtained as the hydrodynamic limit of an Enskog-type kinetic description and we stress, in particular, the role played by spatially non-local interactions among the vehicles towards this result. In Section~\ref{sect:generalisations}, we exploit the Enskog-type hydrodynamics to extend the Aw-Rascle model to a new class of second order macroscopic traffic models, whose characteristic speeds are slower than the mean speed of the flow. We derive these models from a suitable generalisation of the interactions discussed in Section~\ref{sect:micro} and we establish a direct link between the new terms appearing in the macroscopic equations and the features of the new microscopic interaction rules. In Section~\ref{sect:numerics}, we present several numerical experiments, which both validate the theoretical passage from the kinetic to the hydrodynamic descriptions and highlight analogies and differences among the various macroscopic models obtained in the hydrodynamic limit. Finally, in Section~\ref{sect:conclusions}, we present some concluding remarks and we briefly sketch further research prospects.

\section{Microscopic binary interactions}
\label{sect:micro}
One of the leading ideas in kinetic theory is that the important interactions among the particles of the system are \textit{binary}, i.e. each of them involves two particles at a time. Interactions involving simultaneously more than two particles are neglected as higher order effects. In our case, taking inspiration from~\cite{tosin2019MMS}, we express a general binary interaction between any two vehicles as
\begin{equation}
	v'=v+\gamma I(v,\,v_\ast;\,\rho)+D(v)\eta, \qquad v_\ast'=v_\ast,
	\label{eq:binary}
\end{equation}
where $v,\,v_\ast$ and $v',\,v_\ast'$ are the pre- and post-interaction speeds, respectively, of the interacting vehicles. Furthermore, $I$ is a function modelling the speed variation of the $v$-vehicle due to the leading $v_\ast$-vehicle, which, in contrast, does not change speed because of the front-rear anisotropy of the interactions in the traffic stream. We assume that the interaction rule~\eqref{eq:binary} is parametrised by the \textit{traffic density} $\rho$:
$$ \rho(t,\,x):=\int_0^1f(t,\,x,\,v)\,dv, $$
where $f:\R_+\times\R\times [0,\,1]\to\R_+$ is the \textit{kinetic distribution function}, because the global traffic conditions may influence the reactions of the individual drivers. Finally, $\eta$ is a centred random variable, i.e. such that $\ave{\eta}=0$ with $\ave{\cdot}$ denoting expectation, taking into account stochastic fluctuations of the driver behaviour with respect to the deterministic law expressed by $I$. We denote by $\sigma^2>0$ the variance of $\eta$. The function $D$ models the speed-dependent intensity of such a stochastic fluctuation. As far as the variables and the coefficients in~\eqref{eq:binary} are concerned, we will assume
$$ v,\,v_\ast,\,\rho,\,\gamma\in [0,\,1], \quad D(\cdot)\geq 0. $$
In particular, the unitary maximum values of the speed of the vehicles and of the traffic density have to be understood as dimensionless, referred to suitable maximum physical values.

The binary rules~\eqref{eq:binary} do not conserve, either pointwise or on average, the mean speed of the interacting vehicles, indeed
$$ v'+v_\ast'=v+v_\ast+\gamma I(v,\,v_\ast;\,\rho)+D(v)\eta, \qquad
	\ave{v'+v_\ast'}=v+v_\ast+\gamma I(v,\,v_\ast;\,\rho). $$
This is clearly reasonable in view of the physics of vehicle interactions as opposed to that of molecule collisions in classical gas dynamics. Nevertheless, as it is well known in the approach to hydrodynamics by local equilibrium closures, see e.g.~\cite{benedetto1999CMA,benedetto2004CMS,duering2007PHYSA}, in order to obtain a \textit{second order} macroscopic traffic model, namely a model composed of a self-consistent pair of macroscopic equations, it is necessary that the binary interactions~\eqref{eq:binary} conserve locally both the traffic density $\rho$ and the \textit{global mean speed} $u$ defined by
$$ u(t,\,x):=\frac{1}{\rho(t,\,x)}\int_0^1 vf(t,\,x,\,v)\,dv. $$
Indeed, in this way the \textit{local ``Maxwellian''}, i.e. the local equilibrium speed distribution generated by~\eqref{eq:binary}, is parametrised by the conserved quantities $\rho$, $u$, which play the role of the unknowns in the hydrodynamic equations.

The local conservation of $u$ requires a suitable assumption on the binary interaction rule~\eqref{eq:binary}. We recall that if the vehicles are assumed to be homogeneously distributed in space then a statistical description of the superposition of many interactions among them in any point $x$ is provided by the \textit{homogeneous Boltzmann-type equation}, cf.~\cite{pareschi2013BOOK}:
\begin{equation}
	\partial_t\int_0^1\varphi(v)f(t,\,x,\,v)\,dv=	\frac{1}{2}(Q(f,\,f),\,\varphi),
	\label{eq:Boltzmann_hom}
\end{equation}
where $\varphi:[0,\,1]\to\R$ is any test function, also called an \textit{observable quantity}, and $Q$ is the bilinear interaction operator, whose action on a test function $\varphi$ is defined as
$$ (Q(f,\,g),\,\varphi):=\int_0^1\int_0^1\ave{\varphi(v')-\varphi(v)}f(t,\,x,\,v)g(t,\,x,\,v_\ast)\,dv\,dv_\ast. $$
Here, $\ave{\cdot}$ denotes the average with respect to the distribution of the random variable $\eta$ contained in $v'$. Equation~\eqref{eq:Boltzmann_hom} is said to be homogeneous because the space variable plays in it the role of a parameter, so that the statistical description of the traffic is actually the same in every point $x$. Choosing $\varphi(v)=1$, we immediately deduce the local conservation of the traffic density, indeed $\partial_t\rho=0$. With $\varphi(v)=v$ we obtain instead
$$ \partial_tu=\frac{\gamma}{2\rho}\int_0^1\int_0^1 I(v,\,v_\ast;\,\rho)f(t,\,x,\,v)f(t,\,x,\,v_\ast)\,dv\,dv_\ast, $$
therefore $u$ is locally conserved provided
\begin{equation}
	\int_0^1\int_0^1 I(v,\,v_\ast;\,\rho)f(t,\,x,\,v)f(t,\,x,\,v_\ast)\,dv\,dv_\ast=0,
		\quad \forall\,t\in\R_+,\,x\in\R,\,\rho\in [0,\,1].
	\label{eq:cons_u}
\end{equation}
A possible class of functions $I$ satisfying~\eqref{eq:cons_u}, which we will henceforth consider throughout the paper, is
\begin{equation}
	I(v,\,v_\ast;\,\rho):=\Psi(v_\ast;\,\rho)-\Psi(v;\,\rho)
	\label{eq:I}
\end{equation}
for a given $\Psi:[0,\,1]\times [0,\,1]\to\R$.

We conclude this section by observing that, in order to be physically admissible, the binary rules~\eqref{eq:binary} should guarantee $v',\,v_\ast'\in [0,\,1]$ for every choice of $v,\,v_\ast,\,\rho\in [0,\,1]$. This condition is also necessary for the validity of the Boltzmann-type equation in the form~\eqref{eq:Boltzmann_hom}, namely with a constant (unitary, in this case) collision kernel on the right-hand side, which corresponds to considering vehicles as Maxwellian particles. While $v_\ast'\in [0,\,1]$ is obvious, it may be hard to prove, in general, that the same is \textit{a priori} true also for $v'$. Nevertheless, in the simple prototypical case
\begin{equation}
	\Psi(v;\,\rho):=\lambda(\rho)v,
	\label{eq:Psi.v}
\end{equation}
where $\lambda:[0,\,1]\to\R_+$ is a prescribed density-dependent function, it can be proved~\cite{tosin2019MMS} that a sufficient condition for $v'\in [0,\,1]$ is that $\eta$ and $D$ satisfy
\begin{equation}
	\begin{cases}
		\abs{\eta}\leq c(1-\gamma\lambda(\rho)) \\
		cD(v)\leq\min\{v,\,1-v\}
	\end{cases}
	\label{eq:eta.D}
\end{equation}
for an arbitrary constant $c>0$. This implies that $\eta$ is bounded and $D$ vanishes for $v=0,\,1$.

In the particular case $D=0$, a simpler sufficient condition for $v'\in [0,\,1]$ is instead $\lambda(\rho)\leq\frac{1}{\gamma}$.
	
The choice~\eqref{eq:I}-\eqref{eq:Psi.v} leads to the binary interaction
\begin{equation}
	v'=v+\gamma\lambda(\rho)(v_\ast-v)+D(v)\eta, \qquad v_\ast'=v_\ast.
	\label{eq:binary.FTL}
\end{equation}
Apart from the stochastic contribution, by interpreting $\gamma$ as the (small) duration of the interaction we see that the acceleration of the $v$-vehicle, i.e. $\frac{v'-v}{\gamma}$, is proportional to the relative speed with the leading $v_\ast$-vehicle, i.e. $\lambda(\rho)(v_\ast-v)$. This is consistent with the general structure of microscopic follow-the-leader traffic models~\cite{gazis1961OR}, the function $\lambda$ playing the role of the \textit{sensitivity} of the drivers.

\section{Hydrodynamics from a Boltzmann-type description}
\label{sect:hydro.Boltzmann}
A local kinetic description of traffic flow is provided by the following \textit{inhomogeneous} Boltzmann-type equation in weak form:
\begin{equation}
	\partial_t\int_0^1\varphi(v)f(t,\,x,\,v)\,dv+\partial_x\int_0^1 v\varphi(v)f(t,\,x,\,v)\,dv=\frac{1}{2}(Q(f,\,f),\,\varphi),
	\label{eq:Boltzmann_inhom}
\end{equation}
which, by means of the second term on the right-hand side, extends~\eqref{eq:Boltzmann_hom} taking into account also the transport of the vehicles in space according to the kinematic relationship $\dot{x}=v$. See~\cite{prigogine1960OR,prigogine1971BOOK}.

The usual way to derive macroscopic equations for the hydrodynamic parameters, such as $\rho$ and $u$, is to choose $\varphi(v)=v^n$, $n=0,\,1,\,2,\,\dots$, in~\eqref{eq:Boltzmann_inhom}. This procedure is however endless, because the transport term generates systematically a moment of order $n+1$ in the $n$th equation, thereby never making the latter closed. In order to overcome such a difficulty, a typical strategy consists in introducing the following hyperbolic scaling of space and time:
\begin{equation}
	x\to\frac{2}{\varepsilon}x, \qquad t\to\frac{2}{\varepsilon}t,
	\label{eq:hyp_scal}
\end{equation}
with $0<\varepsilon\ll 1$, so that~\eqref{eq:Boltzmann_inhom} becomes\footnote{Also the variables $x,\,t$ of the distribution function $f$ are scaled according to~\eqref{eq:hyp_scal}. However, in order to avoid introducing additional notations, we still denote by $f(t,\,x,\,v)$ the scaled distribution function.}
\begin{equation}
	\partial_t\int_0^1\varphi(v)f(t,\,x,\,v)\,dv+\partial_x\int_0^1 v\varphi(v)f(t,\,x,\,v)\,dv=\frac{1}{\varepsilon}(Q(f,\,f),\,\varphi)
	\label{eq:Boltzmann_inhom.eps}
\end{equation}
In this equation, $\varepsilon$ plays conceptually the role of the Knudsen number of the classical kinetic theory. If $\varepsilon$ is sufficiently small then locally the interactions are much faster than the displacement of the vehicles. As a consequence, a fluid dynamic regime is conceivable, in which the local equilibrium distribution quickly produced by the interactions is simply transported by the traffic stream. This allows one to solve~\eqref{eq:Boltzmann_inhom.eps} by splitting the contributions of the interactions and of the transport:
\begin{align}
	& \partial_t\int_0^1\varphi(v)f(t,\,x,\,v)\,dv=
		\frac{1}{\varepsilon}(Q(f,\,f),\,\varphi) \label{eq:B.split.int} \\
	& \partial_t\int_0^1\varphi(v)f(t,\,x,\,v)\,dv+\partial_x\int_0^1 v\varphi(v)f(t,\,x,\,v)\,dv=0, \label{eq:B.split.transp}
\end{align}
analogously to what is commonly done in the numerical solution of the Boltzmann equation, see e.g.~\cite{dimarco2018JCP,dimarco2014AN,pareschi2001SISC}.

The idea is now that if we are able to identify from~\eqref{eq:B.split.int} the local Maxwellian $M_{\rho,u}$ parametrised by the two conserved quantities $\rho$, $u$, cf.~Section~\ref{sect:micro}, then we may plug it into~\eqref{eq:B.split.transp} to obtain the hydrodynamic equations satisfied by $\rho$, $u$.

\subsection{The case~\texorpdfstring{$\boldsymbol{D\neq 0}$}{}}
\label{sect:B.Dneq0}
Unfortunately, when $D\ne 0$ in~\eqref{eq:binary} it is in general not possible to compute explicitly the steady distributions of the homogeneous Boltzmann-type equation~\eqref{eq:B.split.int}. However, at least in some particular regimes, one may rely on powerful asymptotic procedures, which transform~\eqref{eq:B.split.int} in partial differential equations more amenable to analytical solutions. One of such procedures is the so-called \textit{quasi-invariant interaction limit}, introduced in~\cite{cordier2005JSP} and reminiscent of the \textit{grazing collision limit} applied to the classical Boltzmann equation~\cite{villani1998PhD,villani1998ARMA}.

Let us assume that the system is locally close to equilibrium, so that each binary interaction~\eqref{eq:binary} produces a very small transfer of speed from the leading to the rear vehicle. In particular, we may obtain such an effect by setting
\begin{equation}
	\gamma=\sigma^2=\varepsilon,
	\label{eq:quasi-invariant_scaling}
\end{equation}
which, for $\varepsilon$ small, implies that both the deterministic and the stochastic parts of the interaction are small. In this situation, if $\varphi$ is sufficiently smooth then the difference $\varphi(v')-\varphi(v)$ in~\eqref{eq:B.split.int} can be expanded in Taylor series about $v$. After some computations, this yields
\begin{align*}
	\partial_t\int_0^1\varphi(v)f(t,\,x,\,v)\,dv &= \int_0^1\int_0^1\varphi'(v)I(v,\,v_\ast;\,\rho)f(t,\,x,\,v)f(t,\,x,\,v_\ast)\,dv\,dv_\ast \\
	&\phantom{=} +\frac{1}{2}\int_0^1\varphi''(v)D^2(v)f(t,\,x,\,v)\,dv+R^\varepsilon_\varphi(f,\,f),
\end{align*}
where $R^\varepsilon_\varphi(f,\,f)$ is a bilinear reminder, which, under the assumptions that $I$ is bounded and $\eta$ has bounded third order moment (i.e., $\ave{\abs{\eta}^3}<+\infty$), is asymptotic to $\sqrt{\varepsilon}$ when $\varepsilon\to 0^+$, see~\cite{tosin2019MMS} for the details. On the whole, $R^\varepsilon_\varphi(f,\,f)\to 0$ for $\varepsilon\to 0^+$, so that in such a limit we obtain that $f$ satisfies the equation
\begin{align}
	\begin{aligned}[b]
		\partial_t\int_0^1\varphi(v)f(t,\,x,\,v)\,dv &= \int_0^1\int_0^1\varphi'(v)I(v,\,v_\ast;\,\rho)f(t,\,x,\,v)f(t,\,x,\,v_\ast)\,dv\,dv_\ast \\
		&\phantom{=} +\frac{1}{2}\int_0^1\varphi''(v)D^2(v)f(t,\,x,\,v)\,dv.
	\end{aligned}
	\label{eq:FP.weak}
\end{align}
Integrating by parts the terms on the right-hand side, along with suitable conditions on $f$ at $v=0,\,1$ such that the boundary terms vanish (see again~\cite{tosin2019MMS} for the details), we recognise that this is the weak form of the following \textit{Fokker-Planck equation}:
\begin{equation}
	\partial_tf=\frac{1}{2}\partial^2_v\left(D^2(v)f\right)
		-\partial_v\left(\left(\int_0^1 I(v,\,v_\ast;\,\rho)f(t,\,x,\,v_\ast)\,dv_\ast\right)f\right),
	\label{eq:FP}
\end{equation}
whose solutions approximate the large time behaviour of~\eqref{eq:B.split.int} in the quasi-invariant regime. In particular, the equilibrium solution to~\eqref{eq:FP}, i.e. the local Maxwellian $M_{\rho,u}$, satisfies
$$ \frac{1}{2}\partial_v\left(D^2(v)M_{\rho,u}\right)-\left(\int_0^1 I(v,\,v_\ast;\,\rho)M_{\rho,u}(v_\ast)\,dv_\ast\right)M_{\rho,u}=0, $$
which, for the binary interaction~\eqref{eq:binary.FTL}, cf. also~\eqref{eq:I}-\eqref{eq:Psi.v}, becomes
$$ \frac{1}{2}\partial_v\left(D^2(v)M_{\rho,u}\right)-\lambda(\rho)(u-v)M_{\rho,u}=0, $$
whence
$$ M_{\rho,u}(v)=\frac{C}{D^2(v)}\exp\left(2\lambda(\rho)\int\frac{u-v}{D^2(v)}\,dv\right), $$
$C>0$ being a normalisation constant to be fixed in such a way that $\int_0^1M_{\rho,\,u}(v)\,dv=\rho$.

To proceed further, we have to choose a diffusion coefficient $D$. A closed form of $M_{\rho,u}$ is obtained, for instance, with\footnote{We observe that such a function $D$ does not comply with~\eqref{eq:eta.D}, because of the vertical tangents at $v=0,\,1$. Nevertheless, it can be obtained as the uniform limit, for $\varepsilon\to 0^+$, of a sequence of functions $D_\varepsilon(v)$, which instead satisfy~\eqref{eq:eta.D} for every $\varepsilon>0$, see~\cite{toscani2006CMS}. This justifies its use in the Fokker-Planck equation~\eqref{eq:FP}, i.e. \textit{after} performing the quasi-invariant limit.} $D(v)=\sqrt{v(1-v)}$ and reads
\begin{equation}
	M_{\rho,u}(v)=\rho\frac{v^{2\lambda(\rho)u-1}(1-v)^{2\lambda(\rho)(1-u)-1}}
		{\operatorname{B}(2\lambda(\rho)u,\,2\lambda(\rho)(1-u))},
	\label{eq:M.beta}
\end{equation}
where $\operatorname{B}(\cdot,\,\cdot)$ is the beta function. On the whole, we notice that $\frac{1}{\rho}M_{\rho,u}(v)$ is the probability density function of a beta random variable, interestingly quite consistent with some recent experimental findings about the speed distribution in traffic flow~\cite{maurya2016TRP,ni2018AMM}. Entropy arguments can be invoked~\cite{furioli2017M3AS} to prove that~\eqref{eq:M.beta} is the unique and globally attractive steady solution with mass $\rho$ to~\eqref{eq:FP} with binary rules~\eqref{eq:binary.FTL}.

\begin{remark}
Equation~\eqref{eq:FP.weak} may be rewritten as
$$ \partial_t\int_0^1\varphi(v)f(t,\,x,\,v)\,dv=(\sP(f),\,\varphi), $$
where the \textit{Fokker-Planck operator} $\sP$ is defined, in weak form, as
\begin{align*}
	(\sP(f),\,\varphi) &:= \int_0^1\int_0^1\varphi'(v)I(v,\,v_\ast;\,\rho)f(t,\,x,\,v)f(t,\,x,\,v_\ast)\,dv\,dv_\ast \\
	&\phantom{=} +\frac{1}{2}\int_0^1\varphi''(v)D^2(v)f(t,\,x,\,v)\,dv,
\end{align*}
or equivalently, in strong form, as
$$ \sP(f)(t,\,x,\,v):=\frac{1}{2}\partial^2_v\left(D^2(v)f(t,\,x,\,v)\right)-\partial_v\left(\left(\int_0^1I(v,\,v_\ast;\,\rho)f(t,\,x,\,v_\ast)\,dv_\ast\right)f(t,\,x,\,v)\right). $$

The quasi-invariant limit performed above implies that $Q$ can be \textit{consistently approximated} by $\sP$ in the regime in which $\gamma$, $\sigma^2$ are small and the frequency of the interactions is high.
\label{rem:P.Q}
\end{remark}

Plugging~\eqref{eq:M.beta} into~\eqref{eq:B.split.transp} along with the choices $\varphi(v)=1,\,v$, and recalling the known formulas for the moments of a beta random variable, we obtain the following second order macroscopic model:
\begin{equation}
	\begin{cases}
		\partial_t\rho+\partial_x(\rho u)=0 \\[1mm]
		\partial_t(\rho u)+\partial_x\left(\rho u\dfrac{2\lambda(\rho)u+1}{2\lambda(\rho)+1}\right)=0.
	\end{cases}
	\label{eq:macro.1}
\end{equation}
Introducing the vector of the conserved quantities $U:=(\rho,\,u)^T$, and assuming for simplicity that $\lambda>0$ is constant,~\eqref{eq:macro.1} can be rewritten in quasilinear vector form as
\begin{equation}
\partial_tU+A(U)\partial_xU=0, \qquad
	A(U):=
	\begin{pmatrix}
		u & \rho \\
		\frac{u(1-u)}{(2\lambda+1)\rho} & \frac{(2\lambda-1)u+1}{2\lambda+1}
	\end{pmatrix}.
	\label{eq:Jacobian1}
\end{equation}

In particular, the eigenvalues of $A$ are
$$ \mu_\pm:=u+\frac{1-2u}{2(2\lambda+1)}\pm\frac{1}{2(2\lambda+1)}\sqrt{1+8\lambda u(1-u)}. $$
Since obviously $u\in [0,\,1]$, $\mu_\pm$ are both real, hence system~\eqref{eq:macro.1} is hyperbolic. As it is well known, $\mu_\pm$ represent the speeds of propagation of the small disturbances in the flow and, in macroscopic traffic models, they are required not to exceed the mean speed $u$ of the flow itself. This consistency condition, established in~\cite{aw2000SIAP} in a successful attempt to cure the drawbacks of second order macroscopic traffic models put in evidence in~\cite{daganzo1995TR}, is meant to preserve, at the macroscopic level, the front-rear anisotropy of the microscopic vehicle interactions. We will henceforth call this the \textit{Aw-Rascle condition}.

Unfortunately, for model~\eqref{eq:macro.1} it is immediately evident that
$$ \mu_+\geq u+\frac{1-u}{2\lambda+1}>u \quad \forall\,u\in [0,\,1), $$
thus the hydrodynamic derivation based on the Boltzmann-type local equilibrium closure fails, in general, to produce macroscopic traffic models consistent with the Aw-Rascle condition.

\subsection{The case~\texorpdfstring{$\boldsymbol{D=0}$}{}}
\label{sect:B.Deq0}
If $D=0$ in~\eqref{eq:binary}, i.e. if the stochastic fluctuations ascribable to the driver behaviour are neglected, then it is much easier to compute the Maxwellian $M_{\rho,u}$ directly from~\eqref{eq:B.split.int}. In fact, we see straightforwardly that with $I$ given by~\eqref{eq:I} and for any (continuous) $\Psi$ the distribution
\begin{equation}
	M_{\rho,u}(v)=\rho\delta(v-u),
	\label{eq:M.delta}
\end{equation}
where $\delta$ is the Dirac delta distribution, makes the right-hand side of~\eqref{eq:B.split.int} vanish. Moreover, if $\Psi$ is given in particular by~\eqref{eq:Psi.v} and $0\leq\lambda(\rho)<1$ then from~\eqref{eq:B.split.int} with $\varphi(v)=v^2$ we deduce that the energy of the system converges asymptotically in time to $u^2$ regardless of the initial condition. This implies that~\eqref{eq:M.delta} is the unique and globally attractive steady solution to the interaction step~\eqref{eq:B.split.int}. The Maxwellian~\eqref{eq:M.delta} is also called a \textit{monokinetic distribution}, because it expresses the fact that all vehicles travel locally at the same speed, which coincides with the mean speed of the flow.

Plugging~\eqref{eq:M.delta} into~\eqref{eq:B.split.transp}, we obtain the following second order macroscopic model:
\begin{equation}
	\begin{cases}
		\partial_t\rho+\partial_x(\rho u)=0 \\
		\partial_t(\rho u)+\partial_x(\rho u^2)=0.
	\end{cases}
	\label{eq:macro.2}
\end{equation}
In particular, taking advantage of the first equation, we can rewrite the second equation in the non-conservative form $\partial_tu+u\partial_xu=0$ and, finally, the whole system in quasilinear vector form as
\begin{equation}
	\partial_tU+A(U)\partial_xU=0, \qquad
	A(U):=
	\begin{pmatrix}
		u & \rho \\
		0 & u
	\end{pmatrix}.
	\label{eq:Jacobian2}
\end{equation}
Notice that, due to the monokinetic Maxwellian~\eqref{eq:M.delta}, system~\eqref{eq:macro.2} is \textit{pressureless}. Consequently, the matrix $A(U)$ has two coincident eigenvalues $\mu_\pm=u$, which formally comply with the Aw-Rascle condition. Nevertheless, model~\eqref{eq:macro.2} is much more trivial than the actual Aw-Rascle model~\cite{aw2000SIAP}.

\section{Hydrodynamics from an Enskog-type description}
\label{sect:hydro.Enskog}
The discussion set forth in Section~\ref{sect:hydro.Boltzmann} has shown that, in general, the local equilibrium closure applied to a Boltzmann-type kinetic description of traffic flow fails to yield Aw-Rascle-type hydrodynamic models.  On the other hand, in~\cite{klar1997JSP} the authors already pointed out some inconsistencies in the fluid dynamic behaviour of second order macroscopic traffic models so derived, such as e.g. the inability to reproduce density waves propagating backwards. In particular, they identified the source of such a problem in the \textit{local} nature of the interactions, namely the fact that in~\eqref{eq:Boltzmann_inhom} the two interacting vehicles occupy the same position $x$. Actually, also in the classical Boltzmann equation the colliding gas molecules are supposed to occupy the same space position at the moment of the collision. However, in that case their velocities are not forced to be non-negative, like in the case of the vehicle speeds. This allows one to have, at the macroscopic level, a gas density flowing in principle in any direction.

In order to overcome such difficulties, in~\cite{klar1997JSP} the authors suggested to derive macroscopic traffic models from an \textit{Enskog-type} kinetic description, in which, similarly to the classical Enskog equation for high density gases, the interacting vehicles are \textit{not} supposed to occupy the same position. Specifically, the Enskog-type equation for vehicular traffic takes the form of a modification of the Boltzmann-type equation~\eqref{eq:Boltzmann_inhom}:
\begin{align}
	\begin{aligned}[b]
		\partial_t\int_0^1\varphi(v)f(t,\,x,\,v)\,dv &+ \partial_x\int_0^1 v\varphi(v)f(t,\,x,\,v)\,dv \\
		&= \frac{1}{2}\int_0^1\int_0^1\ave{\varphi(v')-\varphi(v)}f(t,\,x,\,v)f(t,\,x+H,\,v_\ast)\,dv\,dv_\ast,
	\end{aligned}
	\label{eq:Enskog}
\end{align}
where $H>0$ is the headway between the $v$-vehicle and the leading $v_\ast$-vehicle, which here we assume to be constant for simplicity.

If $H$ is small with respect to the characteristic distances along the road, we can write
$$ f(t,\,x+H,\,v_\ast)=f(t,\,x,\,v_\ast)+\partial_xf(t,\,x,\,v_\ast)H+o(H), $$
whence, suppressing the term $o(H)$, we approximate~\eqref{eq:Enskog} as
\begin{align*}
	\partial_t\int_0^1\varphi(v)f(t,\,x,\,v)\,dv &+ \partial_x\int_0^1 v\varphi(v)f(t,\,x,\,v)\,dv \\
	&= \frac{1}{2}\int_0^1\int_0^1\ave{\varphi(v')-\varphi(v)}f(t,\,x,\,v)f(t,\,x,\,v_\ast)\,dv\,dv_\ast \\
	&\phantom{=} +\frac{H}{2}\int_0^1\int_0^1\ave{\varphi(v')-\varphi(v)}f(t,\,x,\,v)\partial_xf(t,\,x,\,v_\ast)\,dv\,dv_\ast.
\end{align*}
From here, performing again the hyperbolic scaling~\eqref{eq:hyp_scal} of space and time, we find
\begin{align}
	\begin{aligned}[b]
		\partial_t\int_0^1\varphi(v)f(t,\,x,\,v)\,dv &+ \partial_x\int_0^1 v\varphi(v)f(t,\,x,\,v)\,dv \\
		&= \frac{1}{\varepsilon}\int_0^1\int_0^1\ave{\varphi(v')-\varphi(v)}f(t,\,x,\,v)f(t,\,x,\,v_\ast)\,dv\,dv_\ast \\
		&\phantom{=} +\frac{H}{2}\int_0^1\int_0^1\ave{\varphi(v')-\varphi(v)}f(t,\,x,\,v)\partial_xf(t,\,x,\,v_\ast)\,dv\,dv_\ast,
	\end{aligned}
	\label{eq:Enskog_approx.eps}
\end{align}
$\varepsilon$ playing again a role analogous to that of the Knudsen number. In particular, if $\varepsilon$ is small we can describe the hydrodynamic regime by means of the following splitting:
\begin{align}
	& \partial_t\int_0^1\varphi(v)f(t,\,x,\,v)\,dv=\frac{1}{\varepsilon}(Q(f,\,f),\,\varphi)
		\label{eq:E.split.int} \\
	& \partial_t\int_0^1\varphi(v)f(t,\,x,\,v)\,dv+\partial_x\int_0^1 v\varphi(v)f(t,\,x,\,v)\,dv=\frac{H}{2}(Q(f,\,\partial_x f),\,\varphi).
		\label{eq:E.split.transp}
\intertext{Notice that~\eqref{eq:E.split.int} is actually the same equation as~\eqref{eq:B.split.int}. In particular, if we consider the regime of small $\gamma$, $\sigma^2$ expressed by the scaling~\eqref{eq:quasi-invariant_scaling} then, in view of Remark~\ref{rem:P.Q}, we can consistently replace (viz. approximate)~\eqref{eq:E.split.int} with}
	& \partial_t\int_0^1\varphi(v)f(t,\,x,\,v)\,dv=(\sP(f),\,\varphi).
		\label{eq:FP.weak.P}
\end{align}
Conversely, unlike~\eqref{eq:B.split.transp}, the transport step~\eqref{eq:E.split.transp} contains a correction on the right-hand side, strictly related to the non-locality of the interactions. With reference to the interaction rules~\eqref{eq:binary.FTL}, we observe that the correction term is such that
\begin{equation}
	(Q(f,\,\partial_xf),\,1)=0, \qquad (Q(f,\,\partial_xf),\,v)=\rho^2\gamma\lambda(\rho)\partial_xu.
	\label{eq:Q.f.df}
\end{equation}

In practice, the idea behind system~\eqref{eq:E.split.int}-\eqref{eq:E.split.transp} may be paraphrased as follows: one determines a local Maxwellian $M_{\rho,u}$ from~\eqref{eq:E.split.int}, as if the interacting vehicles were localised in the same space position. As a matter of fact, this is analytically doable in the quasi-invariant regime, taking advantage of the Fokker-Planck approximation~\eqref{eq:FP.weak.P} of~\eqref{eq:E.split.int}.  Next, one transports $M_{\rho,u}$ by means of~\eqref{eq:E.split.transp}, including a suitable correction to the pure transport~\eqref{eq:B.split.transp} due to the actual non-locality of the interactions. In this transport step, the parameter $\gamma$ appearing on the right-hand side of~\eqref{eq:E.split.transp}, cf.~\eqref{eq:Q.f.df}, will be assumed small, consistently with the quasi-invariant regime invoked to solve~\eqref{eq:E.split.int}.

\subsection{The case~\texorpdfstring{$\boldsymbol{D\neq 0}$}{}}
\label{sect:E.Dneq0}
In the case $D\neq 0$, we can repeat the same steps as in Section~\ref{sect:B.Dneq0} to find the local Maxwellian~\eqref{eq:M.beta}. Plugging it into~\eqref{eq:E.split.transp} with $\varphi(v)=1,\,v$, and recalling furthermore the interaction rule~\eqref{eq:binary.FTL}, we find the following second order macroscopic model:
\begin{equation}
	\begin{cases}
		\partial_t\rho+\partial_x(\rho u)=0 \\[1mm]
		\partial_t(\rho u)+\partial_x\left(\rho u\dfrac{2\lambda(\rho)u+1}{2\lambda(\rho)+1}\right)
			=\rho^2\dfrac{\gamma\lambda(\rho)H}{2}\partial_x u,
	\end{cases}
	\label{eq:macro.3}
\end{equation}
which, assuming again $\lambda>0$ constant for simplicity, can be written in quasilinear vector form as
$$ \partial_tU+A(U)\partial_xU=0, \qquad
	A(U):=
	\begin{pmatrix}
		u & \rho \\
		\frac{u(1-u)}{(2\lambda+1)\rho} & \frac{(2\lambda-1)u+1}{2\lambda+1}-\frac{\gamma\lambda(\lambda-1)H}{2\lambda+1}\rho \\
	\end{pmatrix}, $$
where, as usual, $U:=(\rho,\,u)^T$. Notice that, if $H=0$, both model~\eqref{eq:macro.3} and the matrix $A(U)$ reduce consistently to model~\eqref{eq:macro.1} and matrix $A(U)$ found in Section~\ref{sect:B.Dneq0}.

The eigenvalues of $A(U)$ are, in this case,
$$ \mu_\pm:=\frac{4\lambda u+1}{2(2\lambda+1)}-\frac{\gamma\lambda H\rho}{4}
	\pm\sqrt{\frac{1+8\lambda u(1-u)}{(2(2\lambda+1))^2}+\left(\frac{\gamma\lambda H\rho}{4}\right)^2
		+\frac{\gamma\lambda H\rho}{4(2\lambda+1)}(2u-1)}. $$
Considering that $u\in [0,\,1]$, we estimate:
\begin{align*}
	\mu_+ &\geq \frac{4\lambda u+1}{2(2\lambda+1)}-\frac{\gamma\lambda H\rho}{4}
		+\sqrt{\frac{1}{(2(2\lambda+1))^2}+\left(\frac{\gamma\lambda H\rho}{4}\right)^2
			-\frac{\gamma\lambda H\rho}{4(2\lambda+1)}} \\
	&= \frac{4\lambda u+1}{2(2\lambda+1)}-\frac{\gamma\lambda H\rho}{4}
		+\sqrt{\left(\frac{1}{2(2\lambda+1)}-\frac{\gamma\lambda H\rho}{4}\right)^2} \\
	&= \frac{4\lambda u+1}{2(2\lambda+1)}-\frac{\gamma\lambda H\rho}{4}
		+\abs{\frac{1}{2(2\lambda+1)}-\frac{\gamma\lambda H\rho}{4}}
\intertext{and, assuming $\rho<\frac{2}{(2\lambda+1)\gamma\lambda H}$, we continue this computation as}
	&= \frac{2\lambda u+1}{2\lambda+1}-\frac{\gamma\lambda H\rho}{2}
		=u+\frac{1-u}{2\lambda+1}-\frac{\gamma\lambda H\rho}{2}.
\end{align*}
Finally, if we further restrict ourselves to the case $u<1$, $\rho<\frac{2(1-u)}{(2\lambda+1)\gamma\lambda H}$ (which is a sub-case of the one previously considered), we have $\frac{1-u}{2\lambda+1}-\frac{\gamma\lambda H\rho}{2}>0$, whence we conclude $\mu_+>u$.

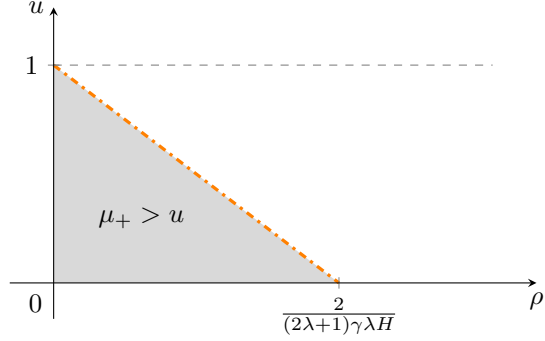
\begin{figure}[!t]
\centering
\resizebox{0.495\textwidth}{!}{
\begin{tikzpicture}[
	declare function={
		f(\x)=1-\x/\a;
	}
]
\pgfmathsetmacro{\a}{1.3}
\begin{axis}[
	axis on top,
	axis lines=middle,
	axis equal image,
	enlarge x limits=0.1,
	enlarge y limits=0.15,
	xtick={0,\a}, xticklabels={0,$\frac{2}{(2\lambda+1)\gamma\lambda H}$},
	ytick={0,1},
	xlabel={$\rho$}, x label style={at={(current axis.right of origin)},anchor=north},
	ylabel={$u$}, y label style={at={(current axis.above origin)},anchor=east},
	xmin=0,
	xmax=2,
	ymin=0,
	ymax=1.1
]
	\coordinate (O) at (0,0);
	\addplot[orange,domain=0:\a,smooth,very thick,style={dash dot},name path=A]{f(x)};
	\addplot[gray,style={dashed}] coordinates {(0,1) (2,1)};
	\path[name path=xaxis] (0,0)--(\a,0);
	\addplot[gray,opacity=0.3] fill between[of=A and xaxis];
	\node at (0.4,0.3){$\mu_+>u$};
	\node at (O) [below left = 1pt of O] {$0$};
\end{axis}
\end{tikzpicture}
}
\caption{The region of the state space $\{(\rho,\,u)\in \R_+\times [0,\,1)\}$ where the greatest eigenvalue of system~\eqref{eq:macro.3} surely exceeds the flow speed $u$. The dash-dotted line has equation $u=1-\frac{(2\lambda+1)\gamma\lambda H}{2}\rho$.}
\label{fig:mu+}
\end{figure}

As Figure~\ref{fig:mu+} shows, the interpretation is that there exists a non-empty subregion of the state space $\{(\rho,\,u)\in\R_+\times [0,\,1)\}$ where $\mu_+$ certainly violates the Aw-Rascle condition. Notice that, for $H\to 0^+$, such a subregion expands to cover the whole state space, consistently with the fact that, as already observed, model~\eqref{eq:macro.3} reduces to model~\eqref{eq:macro.1}.

\subsection{The case~\texorpdfstring{$\boldsymbol{D=0}$}{} and the Aw-Rascle model}
\label{sect:E.Deq0}
If $D=0$ then~\eqref{eq:E.split.int} with the interaction rules~\eqref{eq:binary.FTL} admits again the Maxwellian~\eqref{eq:M.delta} as unique and globally attractive local equilibrium. Plugging it into~\eqref{eq:E.split.transp} yields
\begin{equation}
	\begin{cases}
		\partial_t\rho+\partial_x(\rho u)=0 \\[1mm]
		\partial_t(\rho u)+\partial_x(\rho u^2)=\rho^2\dfrac{\gamma\lambda(\rho)H}{2}\partial_x u.
	\end{cases}
	\label{eq:macro.4}
\end{equation}
Using the first equation, the second equation of this system can be rewritten as
\begin{equation}
	\partial_tu+\left(u-\rho\frac{\gamma\lambda(\rho)H}{2}\right)\partial_xu=0,
	\label{eq:u.AR}
\end{equation}
which coincides with the Aw-Rascle equation for the mean speed upon identifying
\begin{equation}
	p'(\rho):=\frac{\gamma\lambda(\rho)H}{2},
	\label{eq:p'}
\end{equation}
where $p=p(\rho)$ denotes the \textit{traffic ``pressure''}. With this definition,~\eqref{eq:u.AR} can be formally further recast as $\partial_t(u+p(\rho))+u\partial_x(u+p(\rho))=0$, so that finally system~\eqref{eq:macro.4} can be given the usual form of the Aw-Rascle model:
\begin{equation}
	\begin{cases}
		\partial_t\rho+\partial_x(\rho u)=0 \\[1mm]
		\partial_t(u+p(\rho))+u\partial_x(u+p(\rho))=0.
	\end{cases}
	\label{eq:AR}
\end{equation}
In quasilinear vector form this reads
$$ \partial_tU+A(U)\partial_xU=0, \qquad
	A(U):=
		\begin{pmatrix}
			u & \rho \\
			0 & u-\rho p'(\rho)
		\end{pmatrix}, $$
whence we see that the eigenvalues of $A(U)$ are $\mu_-=u-\rho p'(\rho)$, $\mu_+=u$ with clearly $\mu_-\leq u$, because $\lambda(\rho)\geq 0$, hence $p'(\rho)\geq 0$, by assumption (in other words, the traffic pressure $p$ is a non-decreasing function of the traffic density $\rho$).

Summarising, we have been able to recover the Aw-Rascle model organically from first principles of the kinetic theory out of the following \textit{microscopic} features of the binary interactions among the vehicles:
\begin{enumerate}[label=(\roman*)]
\item interactions change only the speed of the vehicles in such a way that the global mean speed is locally conserved;
\item the possible randomness in the behaviour of the drivers is neglected, i.e. driver behaviour is modelled as purely deterministic;
\item interactions are non-local in space, i.e. a headway $H>0$ between the interacting vehicles is taken into account.
\end{enumerate}
Recalling~\eqref{eq:binary.FTL} and~\eqref{eq:p'}, the first two features are realised by means of the interaction rules
$$ v'=v+\dfrac{2}{H}p'(\rho)(v_\ast-v), \qquad v_\ast'=v_\ast. $$
Notice, in particular, that the driver sensitivity $\lambda(\rho)$ turns out to be proportional to the variation of the traffic pressure and inversely proportional to the headway between the interacting vehicles. Thus, the steeper the increase in the traffic pressure, or the closer the leading $v_\ast$-vehicle, the prompter the reaction of the $v$-driver, which is a quite meaningful model of driver behaviour. Moreover, the third feature indicates that Enskog-type equations are the natural kinetic setting for the hydrodynamic derivation of the Aw-Rascle model.

\section{Generalisations of the Aw-Rascle model}
\label{sect:generalisations}
The procedure followed in Section~\ref{sect:E.Deq0} to derive the Aw-Rascle model from the microscopic interactions~\eqref{eq:binary.FTL} can be fruitfully exploited to obtain classes of second order macroscopic traffic models complying with the Aw-Rascle condition.

Let us consider the interaction rules~\eqref{eq:binary} with $I$ given by~\eqref{eq:I} and $D=0$, i.e.
$$ v'=v+\gamma\left(\Psi(v_\ast;\,\rho)-\Psi(v;\,\rho)\right), \qquad v_\ast'=v_\ast. $$
The monokinetic Maxwellian~\eqref{eq:M.delta} is still an equilibrium to~\eqref{eq:E.split.int}, indeed $(Q(M_{\rho,u},\,M_{\rho,u}),\,\varphi)=0$
for every test function $\varphi$. Moreover, considering that $\partial_xM_{\rho,u}=\partial_x\rho\,\delta(v-u)-\rho\partial_xu\,\delta'(v-u)$, we compute
$$ (Q(M_{\rho,u},\,\partial_xM_{\rho,u}),\,1)=0, \qquad
	(Q(M_{\rho,u},\,\partial_xM_{\rho,u}),\,v)=\rho^2\gamma\partial_v\Psi(u;\,\rho)\partial_xu, $$
whence, plugging $M_{\rho,u}$ into~\eqref{eq:E.split.transp} together with $\varphi(v)=1,\,v$, we determine the following macroscopic model:
\begin{equation}
	\begin{cases}
		\partial_t\rho+\partial_x(\rho u)=0 \\[1mm]
		\partial_t(\rho u)+\partial_x(\rho u^2)=\rho^2\dfrac{\gamma\partial_v\Psi(u;\,\rho)H}{2}\partial_xu.
	\end{cases}
	\label{eq:macro.gen}
\end{equation}

Again, using the first equation we can rewrite the second equation in non-conservative form as
$$ \partial_tu+\left(u-\rho\frac{\gamma\partial_v\Psi(u;\,\rho)H}{2}\right)\partial_xu=0, $$
which makes it evident that the quasilinear vector form of system~\eqref{eq:macro.gen} is
$$ \partial_tU+A(U)\partial_xU=0, \qquad
	A(U):=
		\begin{pmatrix}
			u & \rho \\
			0 & u-\rho\frac{\gamma\partial_v\Psi(u;\,\rho)H}{2}
		\end{pmatrix}. $$
The eigenvalues of $A(U)$, i.e. $\mu_-=u-\rho\frac{\gamma\partial_v\Psi(u;\,\rho)H}{2}$ and $\mu_+=u$, satisfy the Aw-Rascle condition provided
\begin{equation}
	\partial_v\Psi(u;\,\rho)\geq 0 \quad \forall\,(\rho,\,u)\in\R_+\times [0,\,1],
	\label{eq:Psi.AR}
\end{equation}
for then it results clearly $\mu_-\leq u$. Under~\eqref{eq:Psi.AR}, we may therefore call~\eqref{eq:macro.gen} a \textit{generalised Aw-Rascle model}. We observe that~\eqref{eq:Psi.AR} requires essentially that $\Psi$ be a non-decreasing function of the speed $v$ for all the physically admissible values of the parameter $\rho$.

Motivated by the introduction of the traffic pressure defined by~\eqref{eq:p'}, which allows one to rewrite the Aw-Rascle model in the form~\eqref{eq:AR}, we introduce now a \textit{generalised traffic pressure} $P=P(\rho,\,u)$ defined by the relationships
\begin{equation}
	\partial_\rho P=\frac{\gamma\partial_v\Psi(u;\,\rho)H}{2}\partial_uP, \qquad P(0,\,u)=u,
	\label{eq:P}
\end{equation}
which allows us to rewrite the generalised Aw-Rascle model~\eqref{eq:macro.gen} in the form
\begin{equation*}
	\begin{cases}
		\partial_t\rho+\partial_x(\rho u)=0 \\[1mm]
		\partial_tP(\rho,\,u)+u\partial_xP(\rho,\,u)=0.
	\end{cases}
\end{equation*}
In practice, $P$ generalises the expression $u+p(\rho)$ in~\eqref{eq:AR}. If, for instance, the function $\Psi$ is such that $\partial_v\Psi(u;\,\rho)$ does not depend on $u$ then from~\eqref{eq:P} we determine precisely $P(\rho,\,u)=u+p(\rho)$ with $p'(\rho):=\frac{\gamma\partial_v\Psi(\rho)H}{2}$, thereby recovering the Aw-Rascle model~\eqref{eq:AR} with $\partial_v\Psi(\rho)=\lambda(\rho)$.

\section{Numerical experiments}
\label{sect:numerics}
In this section, we focus on the numerical description of the models introduced so far. We start from an analysis of the microscopic model of Section~\ref{sect:micro}, which describes the interactions among the vehicles through a binary collision approach. Next, we analyse the various mesoscopic approaches detailed in Sections~\ref{sect:hydro.Boltzmann},~\ref{sect:hydro.Enskog}, investigating in particular the role of the scaling parameter $\varepsilon$. Then, we end with some numerical comparisons between the macroscopic traffic models obtained in the hydrodynamic limit and their corresponding kinetic descriptions. In particular, we show that, for $\varepsilon$ so small that the interactions lead quicly to a local equilibrium, the Enskog model is equivalent to the Aw-Rascle macroscopic model, as anticipated by the theoretical results of Section~\ref{sect:hydro.Enskog}. We also show the anticipating nature of the Enskog model compared to the more standard Boltzmann model.

\subsection{Test 1: Microscopic model and trend to equilibrium}
\label{numerics_I}
We consider the binary rule~\eqref{eq:binary.FTL}, which entails the conservation of both the mass and the global mean speed of the vehicles. In Section~\ref{sect:B.Dneq0}, we have shown that the system reaches a local equilibrium when the number of interactions grows if the effect of each interaction is sufficiently small, i.e. if we are in the so-called quasi-invariant regime. Therefore, we assume, in particular, the quasi-invariant scaling~\eqref{eq:quasi-invariant_scaling} so that, with the choice $D(v)=\sqrt{v(1-v)}$ for the diffusion coefficient, we expect the beta probability density function~\eqref{eq:M.beta} as the local Maxwellian.

Such a Maxwellian depends, on one hand, on the average speed of the vehicles. Since this parameter does not play an important role in the convergence to equilibrium, in the numerical simulations of this section we consider simply a fixed value, specifically $u=0.6$ so as to make the resulting distribution asymmetric. On the other hand, the Maxwellian depends also on the sensitivity parameter $\lambda(\rho)$, whose value strongly affects the shape of the distribution. For the moment, instead of prescribing $\lambda$ as a function of $\rho$, we consider directly several values of $\lambda$, namely $\lambda=1,\,2,\,3,\,4$, and we compare the corresponding Maxwellians emerging from the microscopic dynamics with the analytical expression~\eqref{eq:M.beta} found in the quasi-invariant limit. Moreover, we analyse the convergence of the microscopic model to the equilibrium for two different values of the scaling parameter, specifically $\varepsilon=10^{-1}$ and $\varepsilon=10^{-3}$. As far as the stochastic fluctuation in~\eqref{eq:binary.FTL} is concerned, we consider a uniformly distributed random variable $\eta\sim\mathcal{U}([-0.5,\,0.5])$. As a matter of fact, we notice that the particular type of distribution of $\eta$ does not affect the final Maxwellian but only the transient regime towards it.

\begin{figure}[!t]
\centering
\includegraphics[width=0.75\textwidth]{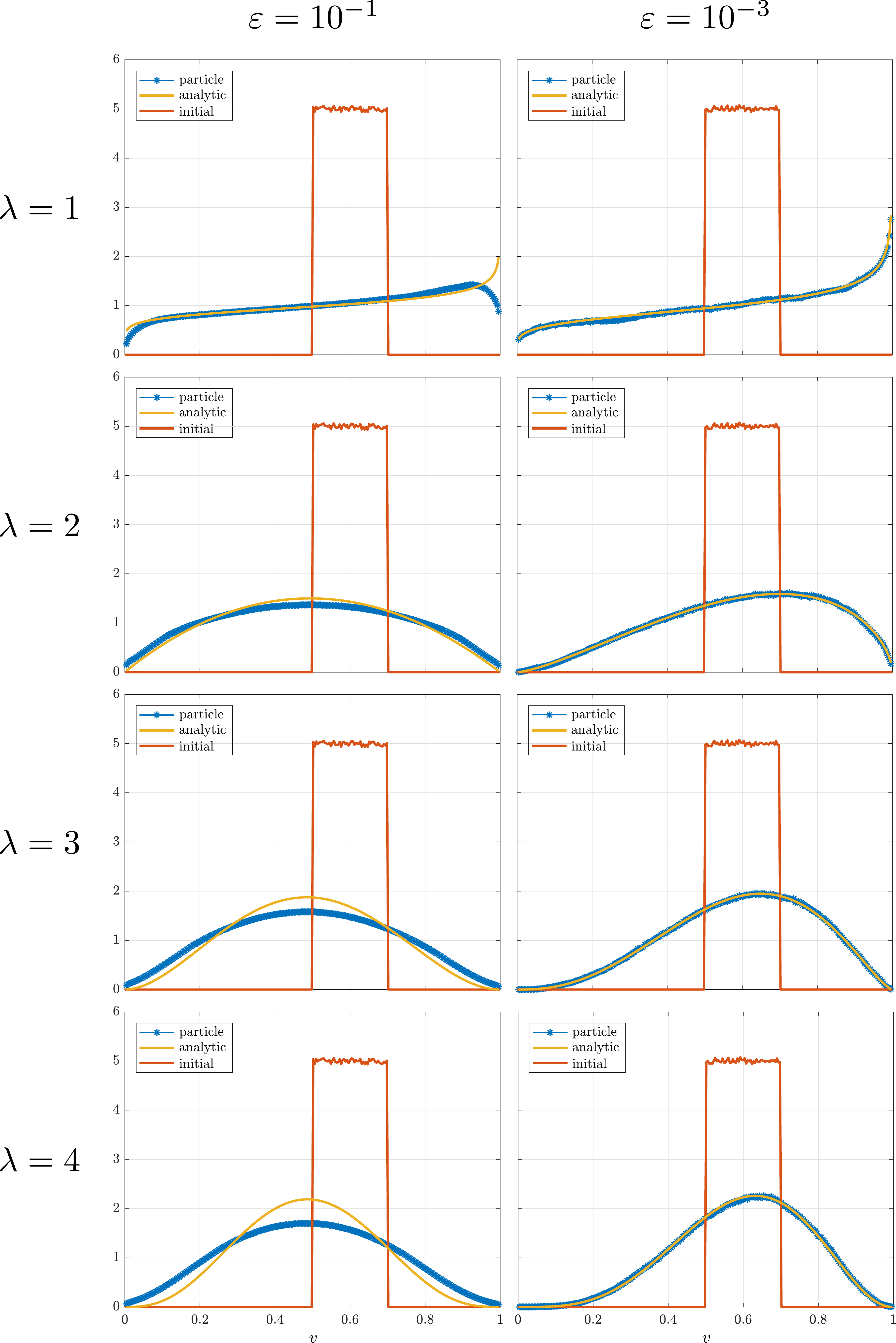}
\caption{\textbf{Test 1}. Different equilibria of the microscopic interaction model~\eqref{eq:binary.FTL} under the quasi-invariant scaling~\eqref{eq:quasi-invariant_scaling}. Left column: $\varepsilon=10^{-1}$, right column: $\varepsilon=10^{-3}$. From top to bottom, the interaction parameter $\lambda$ increases from $1$ to $4$, leading to different shapes of the equilibrium distribution.}
\label{fig:test1}
\end{figure}

In Figure~\ref{fig:test1}, we show the equilibrium distributions obtained with the microscopic dynamics~\eqref{eq:binary.FTL} scaled according to~\eqref{eq:quasi-invariant_scaling}. The curves have been obtained by using $10^3$ vehicles and by averaging the steady state solution over $10^5$ realisations. In each plot, we also represent the analytical steady state~\eqref{eq:M.beta} of the Fokker-Planck equation~\eqref{eq:FP} and the initial distribution $f_0$ of the vehicles, which has been taken uniform in the interval $[0.5,\,2.5]$. In all the tested scenarios, an extremely good agreement between the microscopic interaction dynamics and the Fokker-Planck asymptotics is obtained for $\varepsilon=10^{-3}$, which better mimics the quasi-invariant limit $\varepsilon\to 0^+$.

Finally, we notice that, for $D=0$, the Maxwellian can be determined directly from the Boltzmann-type equation~\eqref{eq:B.split.int} as explained in Section~\ref{sect:B.Deq0}, hence, in particular, without resorting to approximate asymptotic procedures. Such a Maxwellian is therefore exact in every regime of the microscopic parameters and, for this reason, there is no need to report here a numerical comparison.

\subsection{Boltzmann-type model with and without stochasticity}
\label{numerics_II}
We consider now a space non-homogeneous scenario and we report results for the Boltzmann-type description of traffic flow coming from~\eqref{eq:Boltzmann_inhom.eps} under the binary interaction model~\eqref{eq:binary.FTL} along with the quasi-invariant scaling~\eqref{eq:quasi-invariant_scaling}. We start by giving the details of the discretisation technique.

\subsubsection{A Monte Carlo method for the Boltzmann model}
\label{numerics_II_1}
Equation~\eqref{eq:Boltzmann_inhom.eps} is discretised using a Monte Carlo approach, in which we define an ensemble of $N$ particles (representing the vehicles) $\left\{X_k(t),\,V_k(t)\right\}_{k=1}^N$, where $X_k(t)\in\Omega\subseteq\R$ is the position and $V_k(t)\in [0,\,1]$ the speed of the $k$th car at time $t$. Here, $\Omega$ is the space domain. We then approximate the distribution function $f$ by means of the empirical distribution
$$ \mu(t,\,x,\,v):=m_p\sum_{k=1}^{N}\delta(x-X_k(t))\otimes\delta(v-V_k(t)), $$
where the mass $m_p$ of a particle (viz. vehicle) is defined as 
$$ m_p:=\frac{1}{N}\int_\Omega\rho_0(x)\,dx, $$
$\rho_0$ denoting the initial density of the vehicles. Upon introducing in $\Omega\times [0,\,1]$ a space and speed mesh with cell centres $(x_j,\,v_\ell)$ and mesh widths $\Delta{x}$, $\Delta{v}$, respectively, an approximation of the particle density $f(t,\,x,\,v)$ can be obtained as an histogram by computing
\begin{equation}
	f(t,\,x_j,\,v_\ell)=\int_{v_\ell-\Delta{v}/2}^{v_\ell+\Delta{v}/2}\int_{x_j-\Delta{x}/2}^{x_j+\Delta{x}/2}d\mu(t,\,x,\,v).
	\label{eq:hist_RT}
\end{equation}
Likewise, an empirical position density is obtained as
\begin{equation}
	\rho(t,\,x_j)=\int_0^1\int_{x_j-\Delta{x}/2}^{x_j+\Delta{x}/2}d\mu(t,\,x,\,v).
	\label{eq:hist_rho_RT}
\end{equation}

We are now ready to describe the details of the Monte Carlo discretisation. This is based on the strong form of~\eqref{eq:Boltzmann_inhom.eps}, which, by choosing formally $\varphi(\cdot)=\delta(\cdot-v)$, can be written as
\begin{multline}
	\partial_t f(t,\,x,\,v)+v\partial_xf(t,\,x,\,v) \\
	=\frac{1}{\varepsilon}\left[\ave*{\int_0^1\int_0^1\delta(v_1'-v)f(t,\,x,\,v_1)f(t,\,x,\,v_2)\,dv_1\,dv_2}-\rho(t,\,x)f(t,\,x,\,v)\right],
	\label{eq:Boltzmann_inhom_strong}
\end{multline}
where we have denoted by $v_1$, $v_2$ the pre-interaction speeds (dummy integration variables) and by $v_1'$ the post-interaction speed of the first vehicle. The Monte Carlo method corresponding to~\eqref{eq:Boltzmann_inhom_strong} is obtained by splitting the interaction and the transport steps, exactly in the same spirit as~\eqref{eq:B.split.int},~\eqref{eq:B.split.transp}, cf.~\cite{Samaey,pareschi2001ESAIMP}.

\paragraph{Transport}
Each car advances from time $t^n$ over a time interval of length $\Delta{t}$ by changing its position according to
$$ X_k^{n+1}=X_k^n+V_k^n\Delta{t}. $$
This gives an intermediate empirical distribution:
$$ \tilde{\mu}^n(x,\,v):=m_p\sum_{n=1}^{N}\delta(x-X_k^{n+1})\otimes\delta(v-V_k^n), $$
whence the intermediate particle density $\tilde{f}^n(x_j,\,v_\ell)$ and position density $\tilde{\rho}^n(x_j)$ can be computed using~\eqref{eq:hist_RT},~\eqref{eq:hist_rho_RT}.

\paragraph{Interaction}
Next, we solve the interaction step:
$$ \partial_tf(t,\,x,\,v)= \frac{1}{\varepsilon}\left[\ave*{\int_0^1\int_0^1\delta(v_1'-v)f(t,\,x,\,v_1)f(t,\,x,\,v_2)\,dv_1\,dv_2}-\rho(t,\,x)f(t,\,x,\,v)\right], $$
which, by defining the \textit{gain operator}
\begin{equation}
	Q^+(f,\,f)(t,\,x,\,v):=\ave*{\int_0^1\int_0^1\delta(v_1'-v)f(t,\,x,\,v_1)f(t,\,x,\,v_2)\,dv_1\,dv_2},
	\label{eq:Q+}
\end{equation}
can be approximated as
\begin{equation}
	f^{n+1}(x,\,v)=\left(1-\frac{\tilde{\rho}^n(x)\Delta{t}}{\varepsilon}\right)\tilde{f}^{n}(x,\,v)+\frac{\tilde{\rho}^n(x)\Delta{t}}{\varepsilon}Q^+(\tilde{f}^{n},\,\tilde{f}^{n})(x,\,v).
	\label{eq:time_d_coll}
\end{equation}
At the Monte Carlo level,~\eqref{eq:time_d_coll} can be interpreted as follows: 
\begin{itemize}
	\item with probability $1-\frac{\tilde{\rho}^n(x)\Delta t}{\varepsilon}$, the interacting vehicle does not change speed;
	\item with probability $\frac{\tilde{\rho}^n(x)\Delta{t}}{\varepsilon}$, the interacting vehicle changes speed to a new value $V_k^{n+1}$, which is determined by means of Algorithm~\ref{algo1} below.
\end{itemize}
We observe that the explicit time discretisation~\eqref{eq:time_d_coll} requires a stability condition of the type
$$ \max_{x\in\Omega}\frac{\tilde{\rho}^n(x)\Delta{t}}{\varepsilon}\leq 1 $$
in order for the coefficients of the convex combination~\eqref{eq:time_d_coll} to be actual probabilities. This is indeed the choice performed in the numerical results presented in the sequel.

\begin{algorithm}[!t]
	\caption{Nanbu-like algorithm for~\eqref{eq:time_d_coll}}
	\begin{algorithmic}[1]
		\FOR{each cell $j$}
			\STATE define $N_j^n:=$ total number of cars in the cell $j$ at time $t^n=n\Delta{t}$
			\STATE define $N^n_{\text{int},j}:=\left[\dfrac{\tilde{\rho}^n_j\Delta{t}}{\varepsilon}\cdot\dfrac{N_j^n}{2}\right]$, where $[\cdot]$ is a stochastic truncation to the closest integer
			\STATE select uniformly $N^n_{\text{int},j}$ pairs $(k,\,h)$ of vehicles in the cell $j$
			\STATE let the selected pairs interact and, for each of them, set $V_k^{n+1}=V'_k$, $V_h^{n+1}=V'_{h,\ast}$ according to the interaction rule~\eqref{eq:binary.FTL} with the scaling~\eqref{eq:quasi-invariant_scaling}
			\STATE for all the remaining vehicles, set $V_k^{n+1}=V_k^n$
		\ENDFOR
	\end{algorithmic}
	\label{algo1}
\end{algorithm}

The Algorithm~\ref{algo1}, used to compute the post-interaction speed, relies on a so-called Nanbu-type method~\cite{nanbu1983JPSJ}, which is similar to the approach developed for the standard Boltzmann equation of gas dynamics.

\subsubsection{A Finite Volume method for the hydrodynamic limit of the Boltzmann model}
\label{numerics_II_2}
We now detail also the discretisation of the hydrodynamic models~\eqref{eq:macro.1},~\eqref{eq:macro.2}. We use a fifth order WENO method combined with a Rusanov flux for the hyperbolic derivatives~\cite{shu}. Thus, given a generic flux function $F(U)$ with $U\in\R^n$, we first reconstruct the unknown values $U^-$, $U^+$ at the interfaces and then we employ the numerical Rusanov flux defined as:
\begin{align*}
	& H(U^-,\,U^+):=\frac{1}{2}\left[F(U^+)+F(U^-)-\Theta(F')\S(U^+-U^-)\right], \\
	& \Theta(F'):=\max_{U\in[U^-,\,U^+]}\abs{\lambda(F'(U))},
\end{align*}
where $\S\in\R^{n\times n}$ is a transformation matrix and $\max_{U\in[U^-,\,U^+]}\abs{\lambda(F'(U))}$ is the maximum modulus of the eigenvalues of the Jacobian matrix $F'$ of the flux.

For system~\eqref{eq:macro.1}, the Jacobian matrix is given in~\eqref{eq:Jacobian1}. Moreover, we consider $U=(\rho,\,q)^T$ with $q=\rho u$ and the corresponding two components of the flux at the interface:
\begin{align}
	\begin{aligned}[c]
		\hat{f}_{i+\frac{1}{2}} &= \frac{1}{2}\left[f(q^+_{i+\frac{1}{2}})+f(q^-_{i+\frac{1}{2}})-\Theta(\rho,\,q)(\rho^+_{i+\frac{1}{2}}-\rho^-_{i+\frac{1}{2}})\right] \\
		\hat{g}_{i+\frac{1}{2}} &= \frac{1}{2}\left[g(\rho,\,q)^+_{i+\frac{1}{2}}+g(\rho,\,q)^-_{i+\frac{1}{2}}-\Theta(\rho,\,q)(q^+_{i+\frac{1}{2}}-q^-_{i+\frac{1}{2}})\right]
	\end{aligned}
	\label{eq:numflux.1}
\end{align}
with $f(q)=q$, $g(\rho,\,q)=q\frac{2\lambda(\rho)q/\rho+1}{2\lambda(\rho)+1}$ and
$$ \Theta(\rho,\,q)=\frac{q}{\rho}+\frac{1-2q/\rho}{2(2\lambda+1)}\pm\frac{1}{2(2\lambda+1)}\sqrt{1+\frac{8\lambda q}{\rho}\left(1-\frac{q}{\rho}\right)}. $$

For system~\eqref{eq:macro.2}, the Jacobian matrix is given in~\eqref{eq:Jacobian2}. The unknowns $Q=(\rho,\,q)^T$ are the same as before, but the fluxes are different:
\begin{align}
	\begin{aligned}[c]
		\hat{f}^\ast_{i+\frac{1}{2}} &= \frac{1}{2}\left[f^\ast(q ^+_{i+\frac{1}{2}})+f^\ast(q^-_{i+\frac{1}{2}})-\Theta^\ast(\rho,\,q)(\rho^+_{i+\frac{1}{2}}-\rho^-_{i+\frac{1}{2}})\right] \\
		\hat{g}^\ast_{i+\frac{1}{2}} &= \frac{1}{2}\left[g^\ast(\rho,\,q)^+_{i+\frac{1}{2}}+g^\ast(\rho,\,q)^-_{i+\frac{1}{2}}-\Theta^\ast(\rho,\,q)(q^+_{i+\frac{1}{2}}-q^-_{i+\frac{1}{2}})\right]
	\end{aligned}
	\label{eq:numflux.2}
\end{align}
with $f^\ast(q)=q$, $g^\ast(\rho,\,q)=qu$ and $\Theta^\ast(\rho,\,q)=q/\rho$.

The reconstruction of $\rho$, $q$ at the grid interfaces $i\pm 1/2$, necessary for the application of the formulas~\eqref{eq:numflux.1},~\eqref{eq:numflux.2}, may be performed as follows. Let $w$ denote either $\rho$ or $q$. Then the values $w^-_{i+1/2}$, $w^+_{i-1/2}$ are obtained as
$$ w^-_{i+\frac{1}{2}}=\sum_{r=0}^{2}\omega_r w^{(r)}_{i+\frac{1}{2}}, \qquad
	w^+_{i-\frac{1}{2}}=\sum_{r=0}^{2}\omega_r \tilde w^{(r)}_{i+\frac{1}{2}} $$
with weights
$$ \omega_r=\frac{\alpha_r}{\sum_{s=0}^{2}\alpha_s}, \quad \alpha_r=\frac{d_r}{(\epsilon+\beta_r)^2}, \qquad\qquad
	\tilde{\omega}_r=\frac{\tilde{\alpha}_r}{\sum_{s=0}^{2}\tilde{\alpha}_s}, \quad \tilde{\alpha}_r=\frac{\tilde{d}_r}{(\epsilon+\beta_r)^2}, $$
and with the standard smooth indicators
\begin{align*}
	\beta_0 &= \frac{13}{12}(w_i-2w_{i+1}+w_{i+2})^2+\frac{1}{4}(3w_{i}-4w_{i+1}+w_{i+2})^2 \\
	\beta_1 &= \frac{13}{12}(w_{i-1}-2w_{i}+w_{i+1})^2+\frac{1}{4}(w_{i-1}+w_{i+1})^2 \\
	\beta_2 &= \frac{13}{12}(w_{i-2}-2w_{i-1}+w_{i})^2+\frac{1}{4}(3w_{i-2}-4w_{i-1}+w_{i})^2,
\end{align*}
where $\epsilon=10^{-8}$, $d_0=\tilde{d}_2=\frac{3}{10}$, $d_1=\tilde{d}_1=\frac{3}{5}$, $d_2=\tilde{d}_0=\frac{1}{10}$. The values $w^{(r)}_{i\pm\frac{1}{2}}$ represent the third order reconstructions of the pointwise values $\bar{w}_i$. They are obtained through the formulas
$$ w^{(r)}_{i+\frac{1}{2}}=\sum_{j=0}^{2}c_{rj}\bar{w}_{i-r+j}, \qquad w^{(r)}_{i-\frac{1}{2}}=\sum_{j=0}^{2}\tilde{c}_{rj}\bar{w}_{i-r+j}, \qquad r=0,\,1,\,2, $$
where $\bar{w}_{i-r+j}$ are the pointwise values of the unknown evaluated in the points $x_{i-r},\,\dots,\,x_{i-r+2}$. Since we use evenly spaced grid points, the coefficients $c_{rj}$ can be precomputed as indicated in Table~\ref{tab:wenocoef}.

\begin{table}[!t]
	\centering
	\caption{Coefficients $c_{rj}$ for the fifth order space WENO reconstruction on equispaced grid points}\vskip 0.4cm
	\begin{tabular}{|c|rrr|}
		\hline
		\backslashbox{$r$}{$j$} & $0$ & $1$ & $2$ \\
		\hline
		$0$ & $\frac{1}{3}$ & $\frac{5}{6}$ & $-\frac{1}{6}$ \\[2mm]
		$1$ & $-\frac{1}{6}$ & $\frac{5}{6}$ & $\frac{1}{3}$ \\[2mm]
		$2$ & $\frac{1}{3}$ & $-\frac{7}{6}$ & $\frac{11}{6}$ \\
		\hline
	\end{tabular}
	\label{tab:wenocoef}
\end{table}

Finally, we use a second order Runge-Kutta explicit time discretisation. In particular, the time step $\Delta{t}$ is chosen according to the stability condition $\Delta{t}=0.2\Delta{x}/\max_{x\in\Omega}\{\mu_+,\,\mu_-\}$, where $\mu_{\pm}$ are the eigenvalues of the Jacobian matrix of the flux, cf. Sections~\ref{sect:B.Dneq0},~\ref{sect:B.Deq0}.

\subsubsection{Test 2: Boltzmann vs hydrodynamics for~\texorpdfstring{$\boldsymbol{D\neq 0}$}{}}
\label{numerics_II_3}
We now compare the results produced by the inhomogeneous Boltzmann-type model~\eqref{eq:Boltzmann_inhom.eps} for $\varepsilon$ small with those produced by the macroscopic model~\eqref{eq:macro.1} obtained from the former with the local equilibrium closure in the hydrodynamic limit $\varepsilon\to 0^+$.

We consider the space domain $\Omega=[-10,\,10]$ with periodic boundary conditions, which mimics a circuit. To the macroscopic model, we prescribe the following initial condition:
\begin{equation}
	\rho_0(x)=
		\begin{cases}
			0.75 & \text{if } x<0 \\
			0.25 & \text{if } x\geq 0,
		\end{cases}
	\qquad
	u_0(x)=
		\begin{cases}
			0.5 & \text{if } x<0 \\
			0.9 & \text{if } x\geq 0,
		\end{cases}
	\label{eq:test2.macro_initcond}
\end{equation}
which defines a Riemann problem with the discontinuity (shock) located at $x=0$. Notice that, due to the periodic boundary conditions, there is actually also a second discontinuity located at the boundary of $\Omega$, whose left and right states are switched with respect to those of the discontinuity at $x=0$.

\begin{figure}[!t]
\centering
\includegraphics[width=0.75\textwidth]{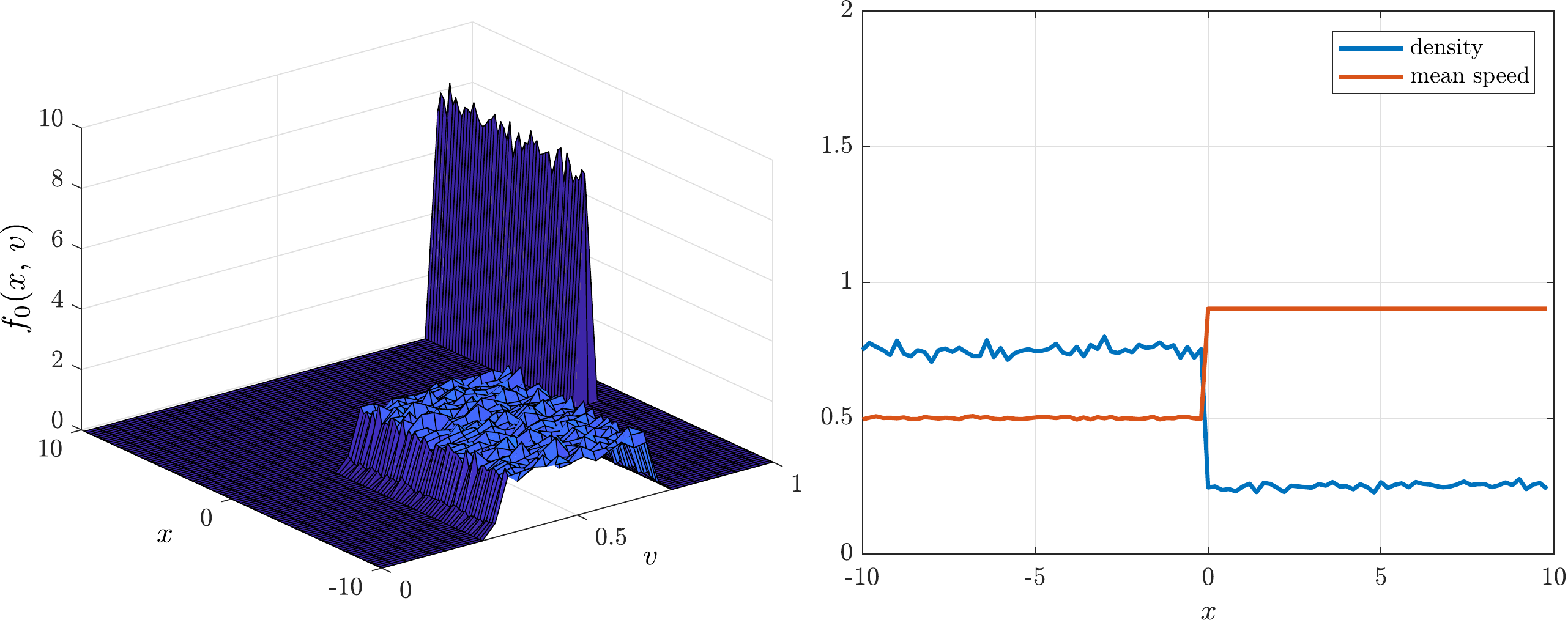}
\caption{\textbf{Test 2}. Left: initial distribution $f_0(x,\,v)$. Right: density and mean speed corresponding to $f_0(x,\,v)$, which mimic consistently the initial condition~\eqref{eq:test2.macro_initcond} of the hydrodynamic model.}
\label{fig:test2_0}
\end{figure}

\begin{figure}[!t]
\centering
\includegraphics[width=0.85\textwidth]{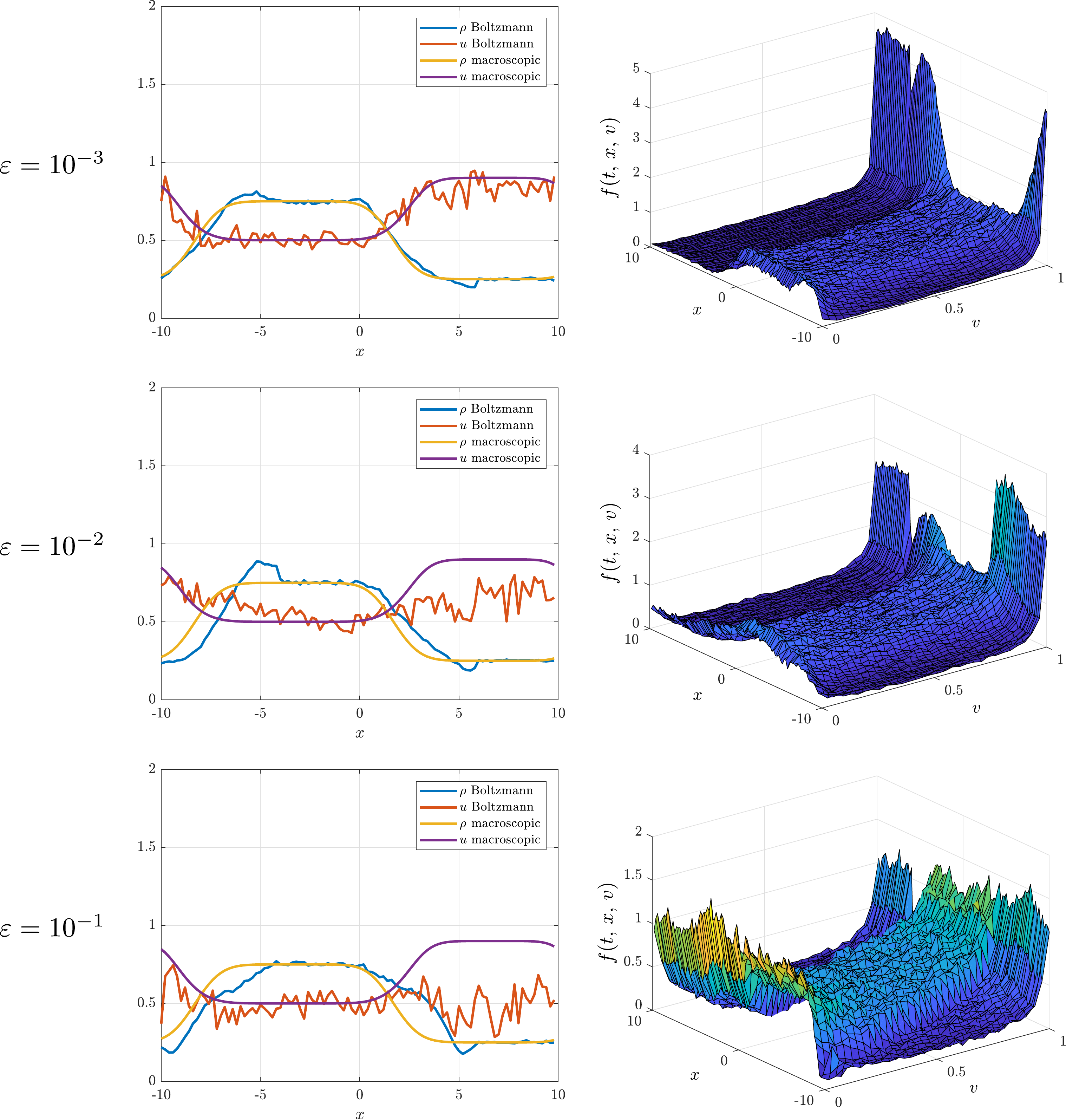}
\caption{\textbf{Test 2}. Left: density and mean speed at the computational time $t=6$ obtained with the hydrodynamic model~\eqref{eq:macro.1} and the Boltzmann-type kinetic model~\eqref{eq:Boltzmann_inhom.eps} with $D\neq 0$ in the binary interactions~\eqref{eq:binary.FTL}. Right: kinetic distribution in the phase space. From top to bottom, the scaling parameter $\varepsilon$ grows from $10^{-3}$ to $10^{-2}$ and $10^{-1}$.}
\label{fig:test2}
\end{figure}

To reproduce such an initial condition at the kinetic level, we use $N=10^4$ particles and we consider a partition of the space domain $\Omega$ in $100$ pairwise disjoint cells. Within each cell, we distribute uniformly a number of vehicles $\rho_0(x_j)/m_p$, $x_j\in\Omega$ being the centre of the $j$th cell, with speed equal to $u_0(x_j)$ plus a small uniform perturbation of the order $p(x_j)u_0(x_j)$, where $p(x_j)=2\cdot 10^{-1}$ if $x_j<0$ and $p(x_j)=10^{-2}$ if $x_j\geq 0$. Hence, we are imposing a non-equilibrium initial condition. Figure~\ref{fig:test2_0} shows the resulting initial distribution $f_0(x,\,v)$ on the left panel and the corresponding density and mean speed, which clearly mimic~\eqref{eq:test2.macro_initcond} consistently, on the right panel.

In Figure~\ref{fig:test2}, we compare the evolution of the system at the computational time $t=6$ calculated via the hydrodynamic and the Boltzmann-type kinetic model. In particular, as far as the latter is concerned, in the microscopic binary interactions~\eqref{eq:binary.FTL} we fix $\lambda(\rho)=\rho$, $\eta\sim\mathcal{U}([-0.5,\,0.5])$ and we use the quasi-invariant scaling~\eqref{eq:quasi-invariant_scaling}. We consider three different orders of magnitude of the scaling parameter: $\varepsilon=10^{-3},\,10^{-2},\,10^{-1}$ (from the top to the bottom of Figure~\ref{fig:test2}). For $\varepsilon=10^{-3}$, we observe a very good matching between the kinetic and the macroscopic solutions, consistently with the fact that the macroscopic model has been obtained from the kinetic model in the limit $\varepsilon\to 0^+$. Conversely, for larger values of $\varepsilon$ some differences appear, because the interactions in the kinetic model are actually farther and farther from the local equilibrium. For instance, for $\varepsilon=10^{-2}$ the maximum mean speed computed with the kinetic model is nearly $0.7$, thus visibly lower than the maximum one computed with the macroscopic model, i.e. $0.9$. This is probably due to diffusive effects, which get more important far from equilibrium. Moreover, we notice that the rarefaction waves characterising the macroscopic solution are slightly shifted rightwards in the kinetic solution. For $\varepsilon=10^{-1}$, we observe even more marked differences with the hydrodynamic solution: the mean speed computed with the kinetic model approaches a nearly constant value around $0.5$ while the waves are much more dampened and shifted rightwards.

\subsubsection{Test 3: Boltzmann vs hydrodynamics for~\texorpdfstring{$\boldsymbol{D=0}$}{}}
\label{numerics_II_4}
Now, we test the case in which the driver behaviour does not contain any stochasticity. Thus, the idea is to repeat the same simulations of Section~\ref{numerics_II_3}, with however $D=0$ in the interaction rules~\eqref{eq:binary.FTL} and still taking the scaling $\gamma=\varepsilon$ into account, cf.~\eqref{eq:quasi-invariant_scaling}. In the limit $\varepsilon\to 0^+$, The expected local equilibrium distribution is the Dirac delta~\eqref{eq:M.delta}, which leads to different dynamics for both the density and the mean speed of the vehicles. Indeed, the obtained hydrodynamic equations are, in this case, the so-called \textit{pressureless gas dynamics equations}, cf.~\eqref{eq:macro.2}.

We prescribe again the initial condition~\eqref{eq:test2.macro_initcond} and we use the same parameters as in the previous test, cf. Section~\ref{numerics_II_3}. Moreover, we solve the Boltzmann-type equation~\eqref{eq:Boltzmann_inhom.eps} by means of the same Monte Carlo method described in Algorithm~\ref{algo1}, cf. Section~\ref{numerics_II_1}, and the hydrodynamics equations~\eqref{eq:macro.2} by means of the same Finite Volume method described in Section~\ref{numerics_II_2}.

\begin{figure}[!t]
\centering
\includegraphics[width=0.85\textwidth]{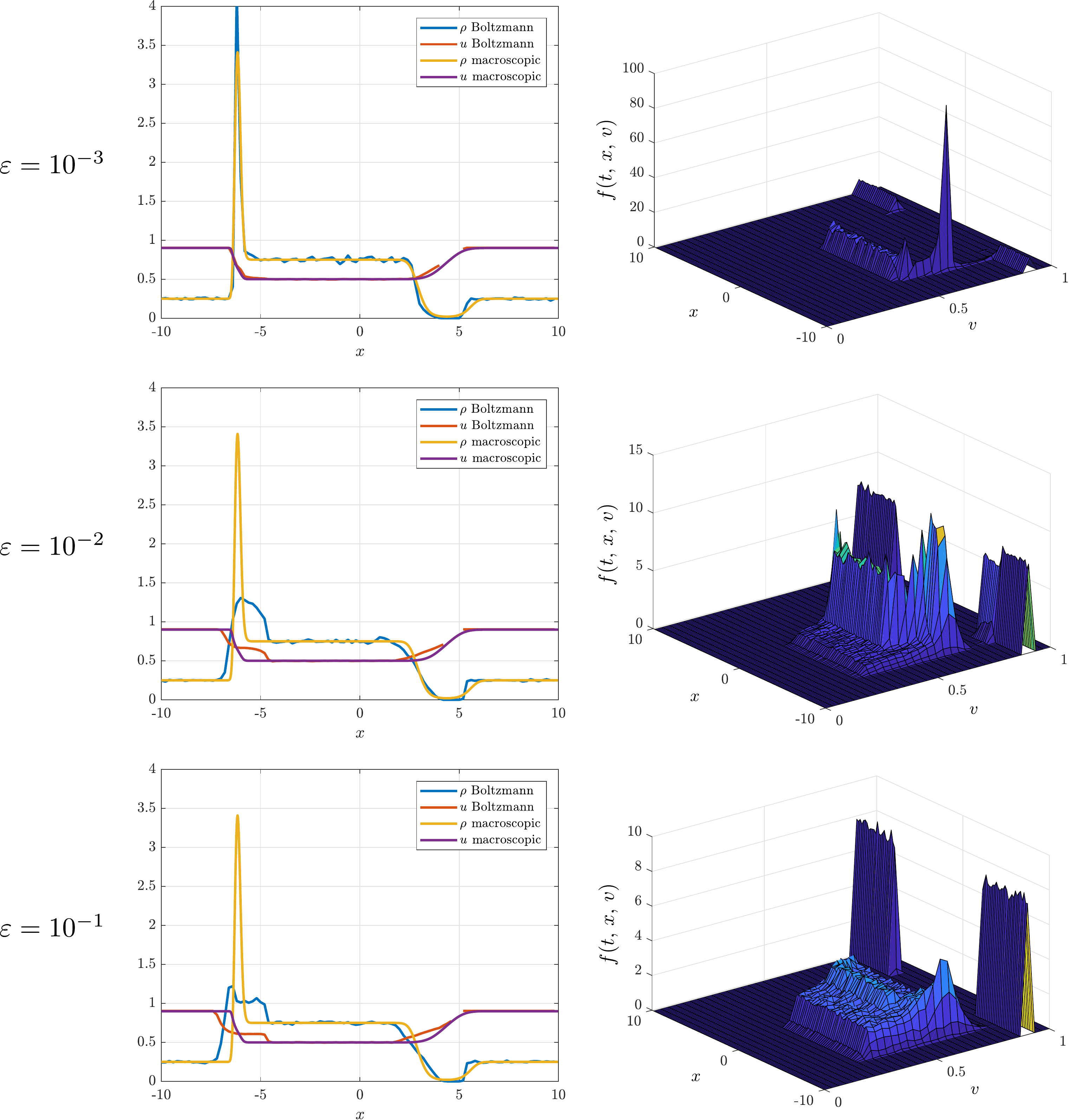}
\caption{\textbf{Test 3}. Left: density and mean speed at the computational time $t=6$ obtained with the hydrodynamic model~\eqref{eq:macro.2} and the Boltzmann-type kinetic model~\eqref{eq:Boltzmann_inhom.eps} with $D=0$ in the binary interactions~\eqref{eq:binary.FTL}. Right: kinetic distribution in the phase space. From top to bottom, the scaling parameter $\varepsilon$ grows from $10^{-3}$ to $10^{-2}$ and $10^{-1}$.}
\label{fig:test3}
\end{figure}

Figure~\ref{fig:test3} shows the results of this test, in particular the density and the mean speed of the vehicles on the left and the kinetic distribution function in the phase space on the right for $\varepsilon=10^{-3},\,10^{-2},\,10^{-1}$ (from top to bottom). The differences with respect to Test 2, cf. Figure~\ref{fig:test2}, are evident by observing the shape of the kinetic distribution function. Compared to the case which includes the stochastic driver behaviour, here $f$ is definitely close to a Dirac delta for $\varepsilon=10^{-3}$. Moreover, also for larger values of $\varepsilon$ the distribution function exhibits important differences with respect to the previous case, although it is not actually concentrated around the mean speed. By analysing the results in terms of the macroscopic parameters, we observe that the density of the vehicles features an incipient vacuum formation near $x=5$, due to a rarefaction caused by the faster vehicles moving rightwards. Such a vacuum formation is well reproduced by the kinetic solution, especially for $\varepsilon=10^{-3}$. Conversely, near $x=-5$ we observe a peak in the density due to faster vehicles reaching the slow traffic region from behind. This makes simultaneously the mean speed decrease because of the congestion. Such a density peak is again very much well reproduced by the kinetic solution for $\varepsilon=10^{-3}$, while for larger values of $\varepsilon$, i.e. far from the hydrodynamic regime, the kinetic solution shows a bump profile with a considerably lower maximum. Parallelly, the mean speed exhibits a much stronger rarefaction wave than in the hydrodynamic case.

\subsection{Enskog-type model with and without stochasticity}
\label{numerics_III}
We pass now to the case of Enskog-type kinetic dynamics. As it is clear from Section~\ref{sect:hydro.Enskog}, in this case we have two different types of interactions: quick local ones, described by~\eqref{eq:E.split.int}, and slower non-local ones modelled by~\eqref{eq:E.split.transp}. In the sequel, we first describe a suitable numerical algorithm consistent with such a splitting of the interaction dynamics, then we compare the kinetic results with the hydrodynamic ones by distinguishing again the cases $D\neq 0$ and $D=0$ in the microscopic interaction rules~\eqref{eq:binary.FTL} with the quasi-invariant scaling~\eqref{eq:quasi-invariant_scaling}.

\subsubsection{A Monte Carlo method for the Enskog model}
\label{numerics_III_1}
In order to derive a Monte Carlo method for the approximation of the Enskog-type model, we first rewrite~\eqref{eq:Enskog_approx.eps} in strong form by choosing formally $\varphi(\cdot)=\delta(\cdot-v)$:
\begin{align}
	\begin{aligned}[b]
		\partial_tf &(t,\,x,\,v)+v\partial_xf(t,\,x,\,v) \\
		&= \frac{1}{\varepsilon}\left[\ave*{\int_0^1\int_0^1\delta(v_1'-v)f(t,\,x,\,v_1)f(t,\,x,\,v_2)\,dv_1\,dv_2}-\rho(t,\,x)f(t,\,x,\,v)\right] \\
		&\phantom{=} +\frac{H}{2}\left[\ave*{\int_0^1\int_0^1\delta(v_1'-v)f(t,\,x,\,v_1)\partial_x f(t,\,x,\,v_2)\,dv_1\,dv_2}-\partial_x\rho(t,\,x)f(t,\,x,\,v)\right],
	\end{aligned}
	\label{eq:Enskog.strong}
\end{align}
where we have again switched to the notation $v_1$, $v_2$ for the pre-interaction speeds (dummy integration variables) and to $v_1'$ for the post-interaction speed of the first vehicle. Next, we choose the size $\Delta{x}$ of the spatial mesh as a submultiple of the headway $H$, i.e. such that $H=k\Delta{x}$ with $k\in\N$. In particular, we fix $k=1$ and we approximate the space derivatives at the right-hand side with the upwind formula:
$$ \partial_x f(t,\,x,\,v_2)\approx\frac{f(t,\,x+\Delta{x},\,v_2)-f(t,\,x,\,v_2)}{\Delta{x}}, \qquad
	\partial_x\rho(t,\,x)\approx\frac{\rho(t,\,x+\Delta{x})-\rho(t,\,x)}{\Delta{x}}. $$
This produces the following approximation of the right-hand side of~\eqref{eq:Enskog.strong}:
\begin{multline}
	\frac{1}{\varepsilon}\left[\ave*{\int_0^1\int_0^1\delta(v_1'-v)f(t,\,x,\,v_1)f(t,\,x,\,v_2)\,dv_1\,dv_2}-\rho(t,\,x)f(t,\,x,\,v)\right] \\
	+\frac{1}{2}\left[\ave*{\int_0^1\int_0^1\delta(v_1'-v)f(t,\,x,\,v_1)f(t,\,x+\Delta{x},\,v_2)\,dv_1\,dv_2}-\rho(t,\,x+\Delta{x})f(t,\,x,\,v)\right],
	\label{eq:Enskog.strong_1}
\end{multline}
where we have further approximated the constant $\frac{1}{\varepsilon}-\frac{1}{2}$ in front of the first term with $\frac{1}{\varepsilon}$, considering that we are interested in the regime of small $\varepsilon$. 

Starting from~\eqref{eq:Enskog.strong_1}, the Monte Carlo method is composed of three steps.

\paragraph{Transport}
Each car advances from time $t^n$ over a time interval of length $\Delta t$ by changing its position according to
$$ X_k^{n+1}=X_k^n+V_k^n\Delta t, $$
whence an intermediate particle density $\tilde{f}^n(x_j,\,v_\ell)$ and the corresponding macroscopic density $\tilde{\rho}^n(x_j)$ can be computed using~\eqref{eq:hist_RT},~\eqref{eq:hist_rho_RT}.

\paragraph{Local interaction}
Invoking the gain operator~\eqref{eq:Q+}, from~\eqref{eq:Enskog.strong_1} with the splitting~\eqref{eq:E.split.int} we update $\tilde{f}^n$ in consequence of the quick local interactions as
\begin{equation}
	\tilde{\tilde{f}}^{n}(x,\,v)=
		\left(1-\frac{\tilde{\rho}^n(x)\Delta{t}}{\varepsilon}\right)\tilde{f}^{n}(x,\,v)
			+\frac{\tilde{\rho}(x)\Delta{t}}{\varepsilon}Q^+(\tilde{f}^{n},\tilde{f}^{n})(x,\,v).
	\label{eq:Enskog.MC_local}
\end{equation}
From $\tilde{\tilde{f}}^n$, we also compute the new macroscopic density $\tilde{\tilde{\rho}}^n$.

\paragraph{Non-local interaction}
Defining from~\eqref{eq:Enskog.strong_1} the Enskog gain operator:
$$ Q^+_E(f,\,f)(t,\,x,\,v):=\ave*{\int_0^1\int_0^1\delta(v_1'-v)f(t,\,x,\,v_1)f(t,\,x+\Delta{x},\,v_2)\,dv_1\,dv_2}, $$
we finally update $\tilde{\tilde{f}}^n$ by taking into account also the contribution of the non-local dynamics:
\begin{equation}
	f^{n+1}(x,\,v)=\left(1-\frac{\tilde{\tilde{\rho}}^n(x+\Delta{x})\Delta{t}}{2}\right)\tilde{\tilde{f}}^{n}(x,\,v)
		+\frac{\tilde{\tilde{\rho}}(x+\Delta{x})\Delta{t}}{2}Q^+_E(\tilde{\tilde{f}}^{n},\,\tilde{\tilde{f}}^{n})(x,\,v).
	\label{eq:Enskog.MC_non-local}
\end{equation}
 
At the Monte Carlo level,~\eqref{eq:Enskog.MC_local}-\eqref{eq:Enskog.MC_non-local} may be interpreted as follows: 
\begin{itemize}
\item in~\eqref{eq:Enskog.MC_local}, with probability $1-\frac{\tilde{\rho}^n(x)\Delta{t}}{\varepsilon}$ the interacting vehicle does not change speed, whereas with probability $\frac{\tilde{\rho}^n(x)\Delta{t}}{\varepsilon}$ it changes speed from $V_k^n$ to a new value $\tilde{V}_k^{n}$ computed by means of~\eqref{eq:binary.FTL} with the quasi-invariant scaling~\eqref{eq:quasi-invariant_scaling}. Such a change of speed is possibly caused by another vehicle located within the same cell of the spatial grid (local interaction);
\item in~\eqref{eq:Enskog.MC_non-local}, with probability $1-\frac{\tilde{\tilde{\rho}}^n(x+\Delta{x})\Delta{t}}{2}$ the interacting vehicle does not change speed, while with probability $\frac{\tilde{\tilde{\rho}}^n(x+\Delta{x})\Delta{t}}{2}$ it changes speed from $\tilde{V}_k^n$ to a new value $V_k^{n+1}$, which is computed again with~\eqref{eq:binary.FTL}-\eqref{eq:quasi-invariant_scaling} but considering now another vehicle located in the next cell of the spatial grid (non-local interaction).
\end{itemize}

The implementation of this algorithm follows very closely the one detailed in Algorithm~\ref{algo1}, thus it is not given here explicitly.

\subsubsection{A Finite Volume method for the hydrodynamic limit of the Enskog model}
\label{numerics_III_1.1}
We now briefly discuss also the numerical approximation of the hydrodynamic equations~\eqref{eq:macro.3},~\eqref{eq:macro.4}.

Both macroscopic models are discretised using a fifth order WENO method combined with a Rusanov flux for the hyperbolic derivatives, as discussed in Section~\ref{numerics_II_2}. In both cases, the unknown is $U=(\rho,\,q)^T$ with $q=\rho u$. Moreover, for~\eqref{eq:macro.3} the numerical flux is given by~\eqref{eq:numflux.1} while for~\eqref{eq:macro.4} it is given by~\eqref{eq:numflux.2} with the same choices of $f$, $g$, $\Theta$, $f^\ast$, $g^\ast$, $\Theta^\ast$ indicated in Section~\ref{numerics_II_2}. Also the reconstruction of the macroscopic parameters $\rho$, $q$ at the grid interfaces follows the same lines outlined in Section~\ref{numerics_II_2}.

The only difference, which requires an \textit{ad hoc} discussion, is the discretisation of the term
$$ \rho^2\frac{\gamma\lambda(\rho)H}{2}\partial_xu $$
appearing on the right-hand side of~\eqref{eq:macro.3},~\eqref{eq:macro.4}, which does not have a counterpart in models~\eqref{eq:macro.1},~\eqref{eq:macro.2} because it is produced by the non-locality of the Enskog interaction operator. We treat this term simply by a time splitting approach. Thus, we first compute
$$ \rho^{n+1/2}_i=\rho^n_i-\frac{\Delta{t}}{\Delta{x}}\left(\hat{f}^n_{i+\frac{1}{2}}-\hat{f}^n_{i-\frac{1}{2}}\right), \qquad
	q^{n+1/2}_i=q^n_i-\frac{\Delta{t}}{\Delta{x}}\left(\hat{g}^n_{i+\frac{1}{2}}-\hat{g}^n_{i-\frac{1}{2}}\right), $$
as if we were solving~\eqref{eq:macro.1},~\eqref{eq:macro.2}, and then we update
$$ \rho^{n+1}_i=\rho^{n+1/2}_i, \qquad
	q^{n+1}_i=q^{n+1/2}_i-\Delta{t}{(\rho^{n+1}_i)}^2\frac{\gamma\lambda(\rho^n_i)H}{2}\cdot\frac{u^{n+1/2}_{i+1}-u^{n+1/2}_i}{\Delta{x}}. $$

Finally, we fix the time step according to the stability condition $\Delta{t}=0.2\Delta{x}/\max_{x\in\Omega}\{\mu_+,\,\mu_-\}$, where $\mu_\pm$ are the eigenvalues of the Jacobian matrix of the flux, cf. Sections~\ref{sect:E.Dneq0},~\ref{sect:E.Deq0}, and we use a second order Runge-Kutta time discretisation.

\subsubsection{Test 4: Enskog vs hydrodynamics for~\texorpdfstring{$\boldsymbol{D\neq 0}$}{}}
\label{numerics_III_2}
\begin{figure}[!t]
\centering
\includegraphics[width=0.85\textwidth]{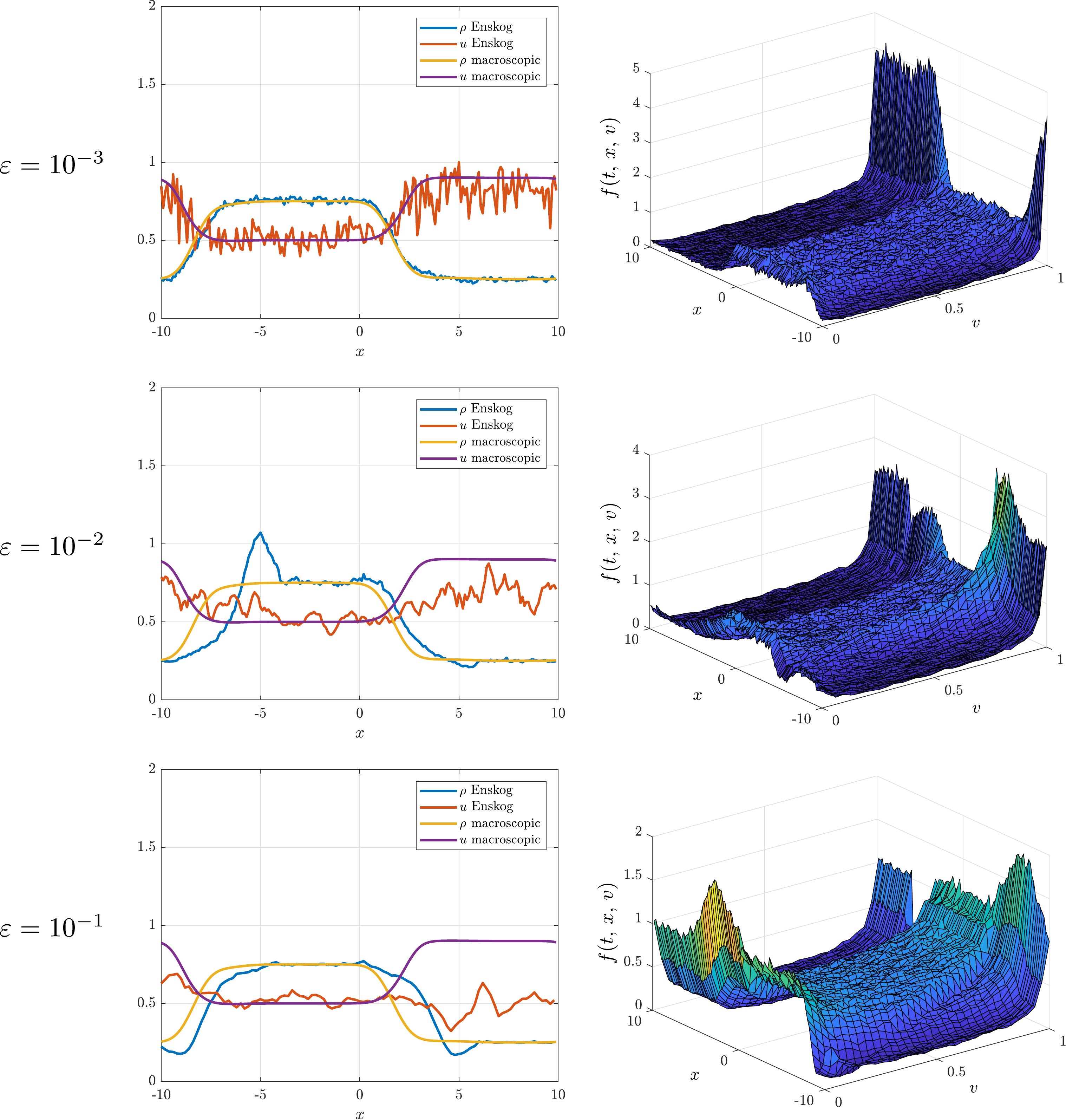}
\caption{\textbf{Test 4}. Left: density and mean speed at the computational time $t=6$ obtained with the hydrodynamic model~\eqref{eq:macro.3} and the Enskog-type kinetic model~\eqref{eq:Enskog_approx.eps} with $D\neq 0$ in the binary interactions~\eqref{eq:binary.FTL}. Right: kinetic distribution in the phase space. From top to bottom, the scaling parameter $\varepsilon$ grows from $10^{-3}$ to $10^{-2}$ and $10^{-1}$.}
\label{fig:test4}
\end{figure}

The setting of this test is the same as that of Section~\ref{numerics_II_3} as far as the details of both the model and the numerical discretisation are concerned, so as to allow for a straightforward comparison. Figure~\ref{fig:test4} shows the results of the present test in terms of the density and the mean speed of the vehicles on the left and of the kinetic distribution function in the phase space on the right for increasing values of the scaling parameter from $\varepsilon=10^{-3}$ to $\varepsilon=10^{-2}$ and $\varepsilon=10^{-1}$ (top to bottom). In particular, the density and the mean speed are computed both from the Enskog-type kinetic model~\eqref{eq:Enskog_approx.eps} and from the corresponding hydrodynamic model~\eqref{eq:macro.3}. As expected from the theory, a good matching between the kinetic and the hydrodynamic solutions is obtained for $\varepsilon=10^{-3}$.

With respect to the Boltzmann-type kinetic model, cf. Figure~\ref{fig:test2}, we observe that the Enskog-type solution exhibits either a more pronounced peak near $x=-5$ for $\varepsilon=10^{-2}$ or a more pronounced incipient vacuum formation near $x=5$ for $\varepsilon=10^{-1}$. On the other hand, for $\varepsilon=10^{-1}$ the mean speed seems to approach again the constant value $0.5$. Instead, the two kinetic solutions, and therefore also the corresponding macroscopic ones, are quite similar to each other near the hydrodynamic regime, namely, in this test, for $\varepsilon=10^{-3}$.

\subsubsection{Test 5: Enskog vs hydrodynamics for~\texorpdfstring{$\boldsymbol{D=0}$}{} (Aw-Rascle model)}
\label{numerics_III_3}
Also this test shares the same modelling and numerical setting as the previous Tests $3$, $4$ discussed in Sections~\ref{numerics_II_3},~\ref{numerics_III_2}. In this case, we compare the Enskog-type kinetic model~\eqref{eq:Enskog_approx.eps} with no microscopic randomness in the driver behaviour ($D=0$ in the binary interactions) and the corresponding hydrodynamic model~\eqref{eq:macro.4}, which, as shown in Section~\ref{sect:E.Deq0}, turns out to be the Aw-Rascle macroscopic traffic model.

\begin{figure}[!t]
\centering
\includegraphics[width=0.85\textwidth]{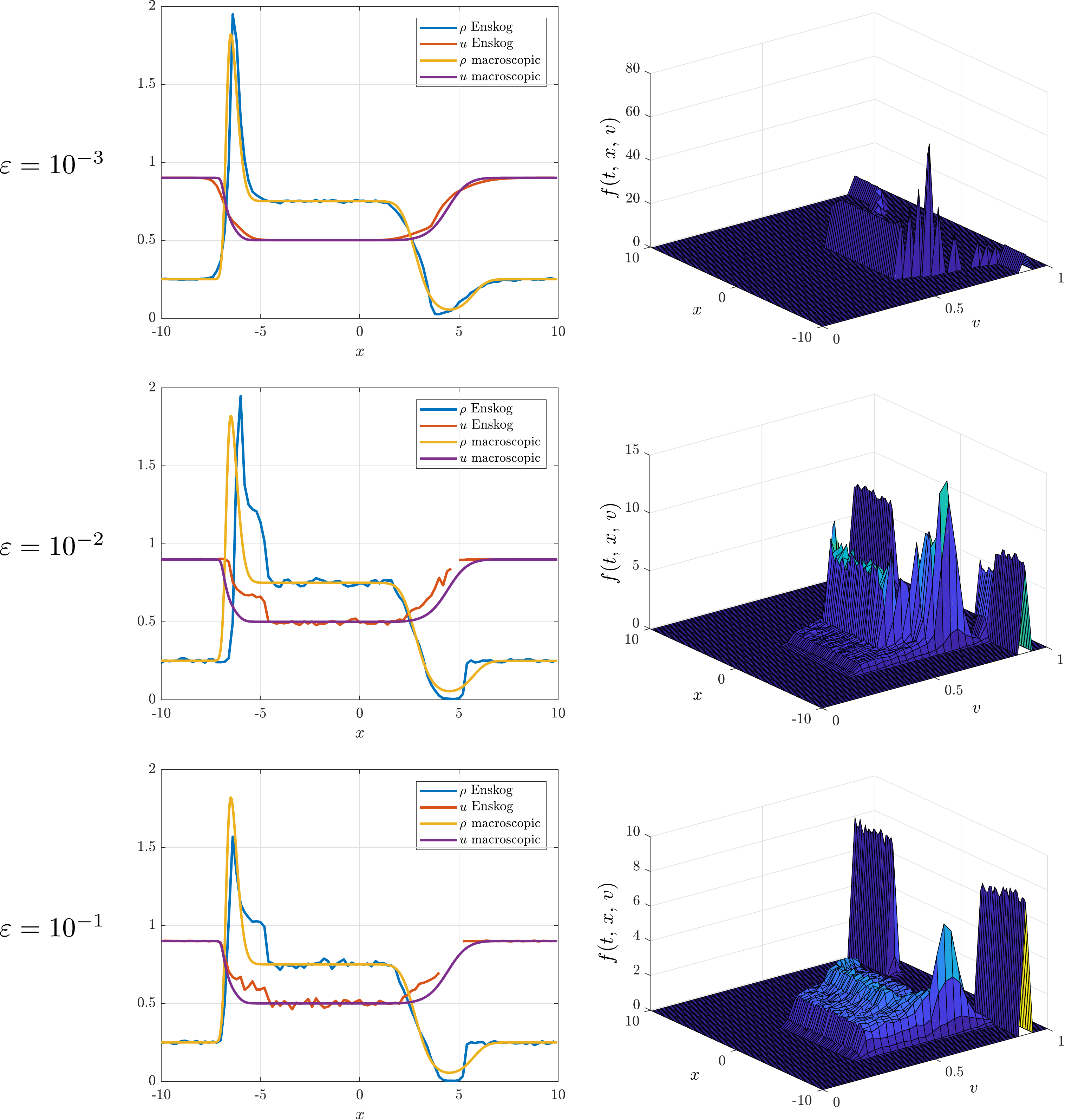}
\caption{\textbf{Test 5}. Left: density and mean speed at the computational time $t=6$ obtained with the hydrodynamic model~\eqref{eq:macro.4} and the Enskog-type kinetic model~\eqref{eq:Enskog_approx.eps} with $D=0$ in the binary interactions~\eqref{eq:binary.FTL}. Right: kinetic distribution in the phase space. From top to bottom, the scaling parameter $\varepsilon$ grows from $10^{-3}$ to $10^{-2}$ and $10^{-1}$.}
\label{fig:test5}
\end{figure}

Figure~\ref{fig:test5} shows the results of this test in terms of the density and the mean speed of the vehicles on the left and of the kinetic distribution function in the phase space on the right for increasing values of the scaling parameter from $\varepsilon=10^{-3}$ to $\varepsilon=10^{-2}$ and $\varepsilon=10^{-1}$ (top to bottom). Compared to the case $D\neq 0$, cf. Section~\ref{numerics_III_2} and Figure~\ref{fig:test4}, the qualitative differences are evident. For small $\varepsilon$, we observe again a very good agreement between the hydrodynamic and the kinetic solutions, as expected from the theory. In particular, the hydrodynamic results are similar to those obtained in the Boltzmann-type case, cf. Figure~\ref{fig:test3}: the rarefaction wave featured by the macroscopic density on the right-hand side of the space domain is stronger, leading to values close to zero, while the peak of the density on the left-hand side of the space domanin gets much more pronounced in the limit $\varepsilon\to 0^+$. For increasing $\varepsilon$, the location of the density peak shifts rightwards in the kinetic solution and the peak value diminishes.

\subsection{Comparison of the hydrodynamic models}
\label{numerics_IV}
Finally, we compare the macroscopic models~\eqref{eq:macro.1},~\eqref{eq:macro.2},~\eqref{eq:macro.3},~\eqref{eq:macro.4} obtained in the hydrodynamic limit. The numerical discretisation is the same as that described in Sections~\ref{numerics_II_2},~\ref{numerics_III_1.1}, however here we use a finer spatial grid made of $500$ cells in order to better highlight the differences among the various cases.

\begin{figure}[!t]
\centering
\includegraphics[width=0.9\textwidth]{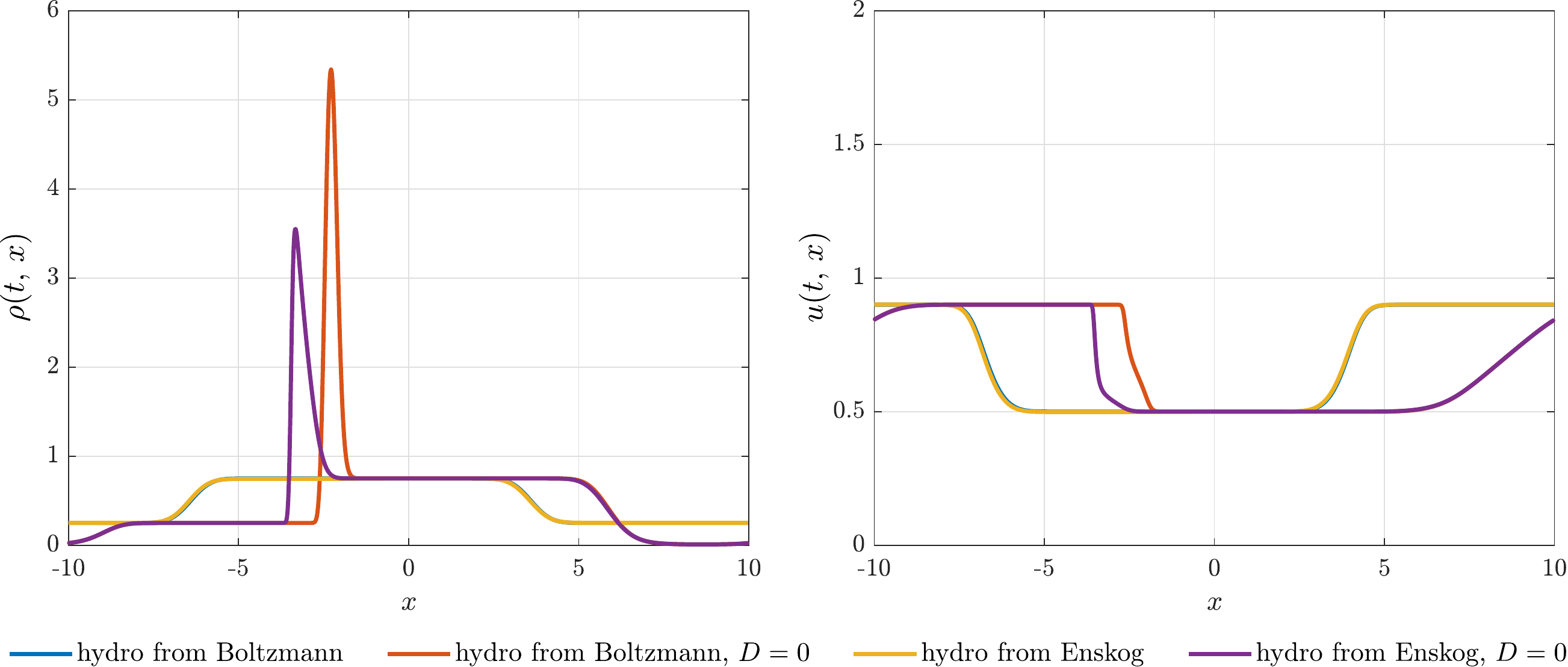}
\caption{\textbf{Test 6}. Density $\rho$ (left) and mean speed $u$ (right), both at the computational time $t=12$, computed with the four hydrodynamic models~\eqref{eq:macro.1},~\eqref{eq:macro.2},~\eqref{eq:macro.3},~\eqref{eq:macro.4}, starting from the initial condition~\eqref{eq:test2.macro_initcond} and with periodic boundary conditions.}
\label{fig:test6}
\end{figure}

\paragraph{Test 6}
We begin from the same Riemann problem that we have considered so far, namely the one described by the initial conditions~\eqref{eq:test2.macro_initcond}. In Figure~\ref{fig:test6}, we report the macroscopic density and the mean speed produced by the four models at the computational time $t=12$. We clearly observe that the hydrodynamic models~\eqref{eq:macro.2},~\eqref{eq:macro.4}, obtained from the Boltzmann-type and the Enskog-type kinetic descriptions with $D=0$, respectively, show the most relevant differences. In particular, the solution of model~\eqref{eq:macro.4} is shifted leftwards with respect to that of model~\eqref{eq:macro.2} and, furthermore, it produces milder congestion states due to the anticipatory ability of the drivers. Conversely, the hydrodynamic models~\eqref{eq:macro.1},~\eqref{eq:macro.3}, obtained from the Boltzmann-type and the Enskog-type kinetic descriptions with $D\neq 0$, respectively, feature very little differences from each other, due to the smoothing role played by the microscopic diffusion.

\paragraph{Test 7}
We now consider the propagation of an initially smooth wave, that we choose as
\begin{equation}
	\rho_0(x)=\frac{1}{3}\left[2+\sin\left(\frac{\pi}{5}x\right)\right], \qquad
		u_0(x)=\frac{1}{2+\sin\left(\frac{\pi}{5}x\right)}.
	\label{eq:test7.macro_initcond}
\end{equation}

\begin{figure}[!t]
\centering
\includegraphics[width=0.9\textwidth]{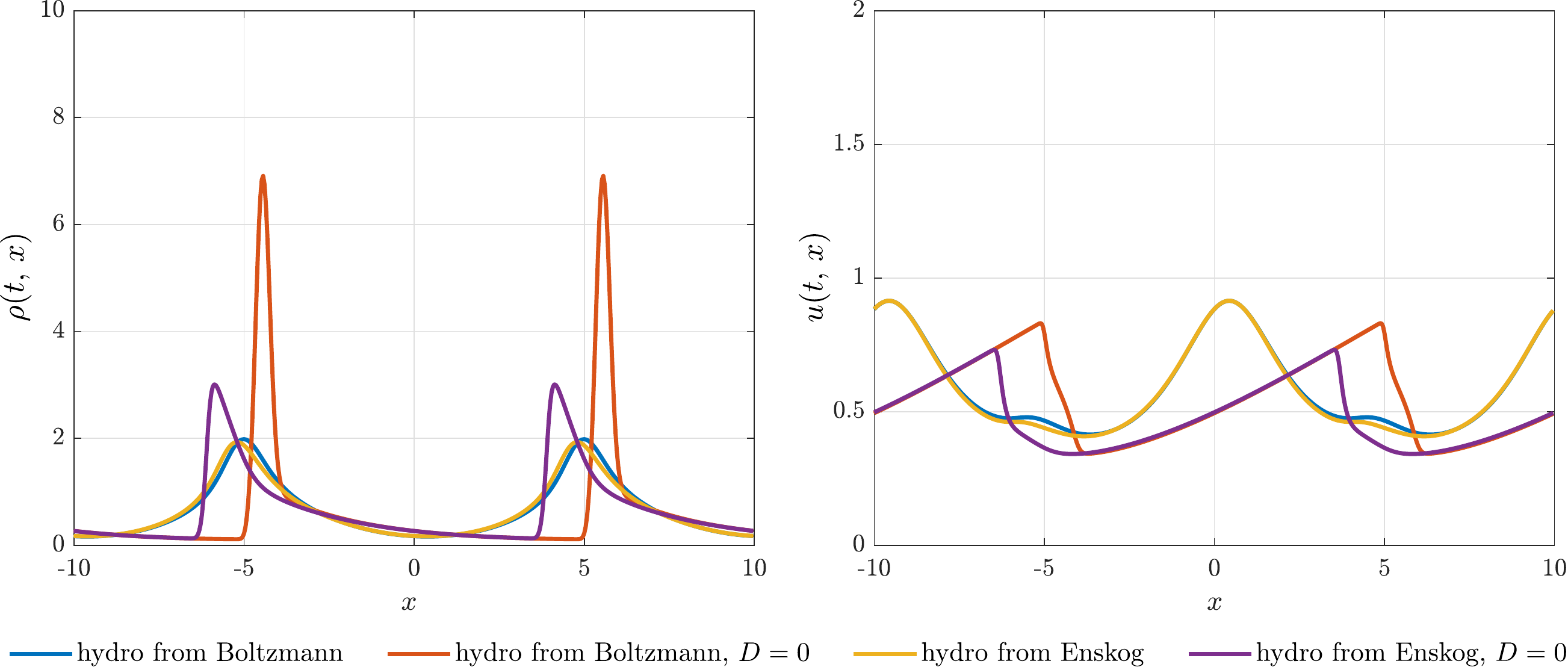}
\caption{\textbf{Test 7}. Density $\rho$ (left) and mean speed $u$ (right), both at the computational time $t=12$, computed with the four hydrodynamic models~\eqref{eq:macro.1},~\eqref{eq:macro.2},~\eqref{eq:macro.3},~\eqref{eq:macro.4}, starting from the initial condition~\eqref{eq:test7.macro_initcond} and with periodic boundary conditions.}
\label{fig:test7}
\end{figure}

Figure~\ref{fig:test7} shows that the results are qualitatively analogous to those of the previous test: the two hydrodynamic models with $D\neq 0$, namely~\eqref{eq:macro.1} and~\eqref{eq:macro.3}, do not exhibit important differences, which are instead more marked in the case $D=0$ between the hydrodynamic models~\eqref{eq:macro.2} and~\eqref{eq:macro.4}. For $D=0$, a density congestion arises. More specifically, in model~\eqref{eq:macro.2}, obtained from the Boltzmann-type description, the congestion is stronger, while in model~\eqref{eq:macro.4}, obtained from the Enskog-type description, it is milder due to the anticipatory nature of the interactions among the vehicles, which start to slow down before reaching the queue. Conversely, for $D\neq 0$ we observe a regularising effect of the microscopic diffusion on the macroscopic traffic, together with wave propagation phenomena similar to the case of a linear hyperbolic system.

\paragraph{Test 8}
Finally, we increase the strength of the interaction with the vehicles ahead, so as to ascertain the effect of the anticipatory nature of the interactions in the Enskog-type setting. In this test, we confine ourselves to the case $D=0$, namely we compare the pressureless hydrodynamic model~\eqref{eq:macro.2} with the Aw-Rascle hydrodynamic model~\eqref{eq:macro.4}. In particular, we set $\lambda(\rho)=10\rho$, while we remind that, so far, we have always used $\lambda(\rho)=\rho$. All of the other parameters are set as in the previous tests. Furthermore, we prescribe the following initial condition:
\begin{equation}
	\rho_0(x)=
		\begin{cases}
			0.25 & \text{if } x<0 \\
			0.75 & \text{if } x\geq 0,
		\end{cases}
	\qquad
	u_0(x)=
		\begin{cases}
			0.2 & \text{if } x<0 \\
			0.4 & \text{if } x\geq 0
		\end{cases}
	\label{eq:test8.macro_initcond}
\end{equation}
and we assign periodic boundary conditions. Similarly to~\eqref{eq:test2.macro_initcond}, this setting defines a Riemann problem with two discontinuities located in $x=0$ and at the boundary of the domain, respectively.

\begin{figure}[!t]
\centering
\includegraphics[width=0.85\textwidth]{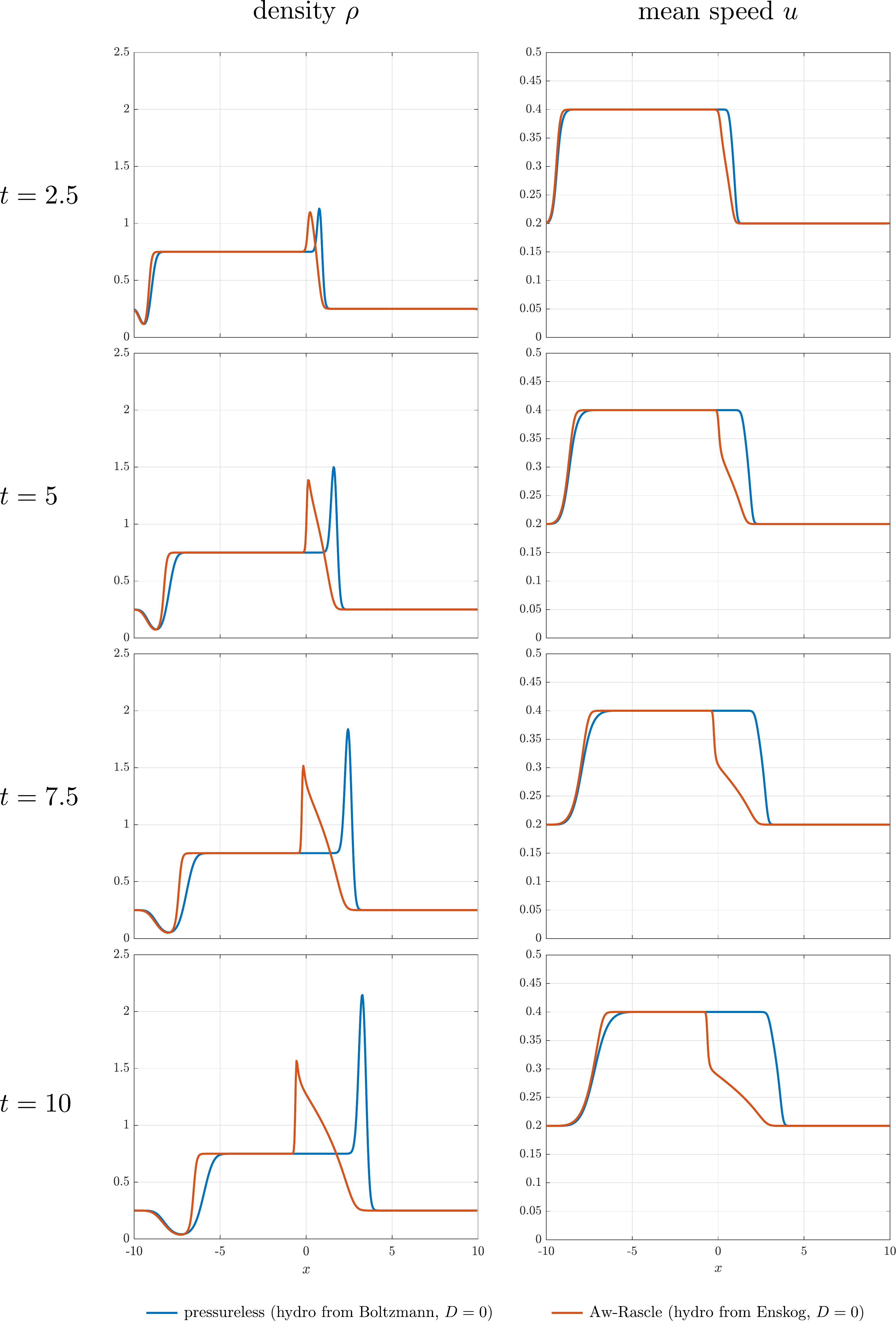}
\caption{\textbf{Test 8}. Density (left) and mean speed (right) at successive times computed with the hydrodynamic models~\eqref{eq:macro.2},~\eqref{eq:macro.4} starting from the initial condition~\eqref{eq:test8.macro_initcond} and with periodic boundary conditions.}
\label{fig:test8}
\end{figure}

Figure~\ref{fig:test8} shows the evolution of the density and of the mean speed, respectively, at the successive computational times $t=2.5,\,5,\,7.5,\,10$. We clearly see that the Aw-Rascle model~\eqref{eq:macro.4}, obtained from the Enskog-type description with $D=0$, is able to anticipate the situation of the traffic ahead and to account for a backward propagation of density waves. This also allows the model to produce bounded congestion states. Conversely, the pressureless model~\eqref{eq:macro.2}, obtained from the Boltzmann-type description with $D=0$, forms a stronger and stronger localised congestion, which moves forward. At the same time, in the Aw-Rascle model the mean speed is lower than that of the pressureless model in correspondence of the traffic congestion.

The same qualitative results, here investigated in the limit hydrodynamic models, hold for the Boltzmann-type and the Enskog-type kinetic models with a small enough scaling parameter $\varepsilon>0$. Consistently with what was argued in~\cite{klar1997JSP}, this indicates that, in spite of the non-negativity of the microscopic speeds, the Enskog-type description, unlike the Boltzmann-type one, is able to account for backward propagating density waves. Not surprisingly, then, it constitutes an appropriate basis for the derivation of the hydrodynamic Aw-Rascle model from kinetic principles.

\section{Conclusions}
\label{sect:conclusions}
In this paper, we have proved that the macroscopic Aw-Rascle traffic model~\cite{aw2000SIAP}, proposed independently also by Zhang~\cite{zhang2002TRB}, may be fruitfully explained as the hydrodynamic limit of an Enskog-type kinetic description. In particular, we have shown that the non-locality of the microscopic interactions among the vehicles plays a fundamental role in conferring the anticipatory nature on the macroscopic dynamics. More precisely, our results indicate that such a large scale anticipatory behaviour is produced by the superposition of quick local interactions and slow background actions, that the drivers perform to adapt their pace to the mean flow. Background actions are not present in a more standard Boltzmann-type kinetic description, which explains why, as already implied in~\cite{klar1997JSP}, the Aw-Rascle model cannot be obtained therefrom. However, our study has highlighted that the non-locality of the interactions is not sufficient, by itself, to produce, in the macroscopic limit, the Aw-Rascle model. It is also necessary that the microscopic rules followed by the vehicles, here written as a relaxation towards the speed of the leading vehicle (follow-the-leader), are deterministic, i.e. free from stochastic contributions caused by the driver behaviour. Indeed, in this way the local interactions drive the system quickly towards the local mean speed of the flow without fluctuations. If present, instead, the random fluctuations generate a standard gas-dynamical pressure-like term in the macroscopic equations, which, as explained in~\cite{aw2000SIAP}, is typically responsible for the violation of the Aw-Rascle condition.

By relying on this sound mathematical-physical understanding, we have used the Enskog-type kinetic description and the related hydrodynamic limit to generalise the Aw-Rascle model to a new class of second order macroscopic traffic models, which satisfy the Aw-Rascle condition. Also for such new models, it has been possible to link precisely the key macroscopic features, such as e.g., the Aw-Rascle traffic pressure responsible for the anticipatory dynamics, to structural properties of the microscopic binary rules modelling the behaviour of the vehicles.

We believe that the link we have established in this paper between the kinetic approach and possibly generalised versions of the Aw-Rascle model may pave the way for the future investigation of hierarchical control problems, from the level of single vehicles (driver-assist/autonomous vehicles) to that of the aggregate flow, in the spirit of~\cite{tosin2019MMS} and with specific focus on second order macroscopic traffic models.

\section*{Acknowledgements}
This research was partially supported  by the Italian Ministry for Education, University and Research (MIUR) through the ``Dipartimenti di Eccellenza'' Programme (2018-2022), Department of Mathematical Sciences ``G. L. Lagrange'', Politecnico di Torino (CUP: E11G18000350001) and through the PRIN 2017 project (No. 2017KKJP4X) ``Innovative numerical methods for evolutionary partial differential equations and applications''.

This work is also part of the activities of the Starting Grant ``Attracting Excellent Professors'' funded by ``Compagnia di San Paolo'' (Torino) and promoted by Politecnico di Torino.

G.D., A.T. are respectively members of GNCS (Gruppo Nazionale per il Calcolo Scientifico) and of GNFM (Gruppo Nazionale per la Fisica Matematica) of INdAM (Istituto Nazionale di Alta Matematica), Italy.

\bibliographystyle{plain}
\bibliography{DgTa-Aw-Rascle_Enskog}
\end{document}